\documentclass[12pt]{article}

\advance\hoffset by -7mm   
\makeatletter 
\@addtoreset{equation}{section}  
\makeatother 
 
\def\ftoday{{\sl {Le \number\day \space\ifcase\month  
\or janvier\or f\'evrier\or mars\or avril\or mai 
\or juin\or juillet\or ao\^ut\or septembre\or octobre 
\or novembre \or d\'ecembre\fi\space \number\year}}}     
\def\ptoday{{\sl {\number\day \space de\space \ifcase\month  
\or janeiro\or fevereiro\or mar{\c c}o\or abril\or maio 
\or junho\or julho\or agosto\or setembro\or outubro 
\or novembro \or dezembro\fi\space de\space \number\year}}}     
\def\gtoday{{\sl {Den \number\day. \ifcase\month  
\or Januar\or Februar\or M\"arz\or April\or Mai 
\or Juni\or Juli\or August\or September\or Oktober 
\or November \or Dezember\fi\space \number\year}}}     
\def\today{{\sl {\ifcase\month 
\or January\or February\or March\or April\or May 
\or June\or July\or August\or September\or October 
\or November \or December\fi \space\number\day,\space  
                                            \number\year}}} 
           \newcommand{\G}{\Gamma} 
         \newcommand{\D}{\Delta} 
\newcommand{\e}{\varepsilon}

\newcommand{\m}{\mu} 
\newcommand{\n}{\nu} 
\newcommand{\om}{\omega}         \newcommand{\OM}{\Omega} 
\newcommand{\p}{\psi}               
\newcommand{\s}{\sigma}            
\renewcommand{\th}{\theta}          
\newcommand{\f}{{\phi}}            
\newcommand{\vf}{{\varphi}} 
\renewcommand{\AA}{{\cal A}}

\newcommand{\EE}{{\cal E}} 
\newcommand{\FF}{{\cal F}}

\newcommand{\LL}{{\cal L}} 
\newcommand{\MM}{{\cal M}} 
 
\newcommand{\OO}{{\cal O}}

\newcommand{\RR}{{\cal R}} 
\newcommand{\TT}{{\cal T}}

\newcommand{\cas}{{\mbox{\footnotesize$\cal S$}}}
\newcommand{\es}{\\[1.7mm]}

\newcommand{\sla}{\raise.15ex\hbox{$/$}\kern -.57em}  
\newcommand{\Sla}{\raise.15ex\hbox{$/$}\kern -.70em}

\newcommand{\lp}{\left(}\newcommand{\rp}{\right)} 
\newcommand{\lc}{\left[}\newcommand{\rc}{\right]}

\newcommand{\complex}{{\kern .1em {\raise .47ex 
\hbox {$\scriptscriptstyle |$}} 
    \kern -.4em {\rm C}}} 
\newcommand{\real}{{{\rm I} \kern -.19em {\rm R}}} 
\newcommand{\rational}{{\kern .1em {\raise .47ex 
\hbox{$\scripscriptstyle |$}} 
    \kern -.35em {\rm Q}}} 
\renewcommand{\natural}{{\vrule height 1.6ex width 
.05em depth 0ex \kern -.35em {\rm N}}}

\newcommand{\tr}{{\rm {Tr} \,}} 
 
\newcommand{\pa}{\partial}

\newcommand{\dsum}[2]{\displaystyle{\sum_{#1}^{#2}}}

\newcommand{\twiddle}{\lower.9ex\rlap{$\kern -.1em\scriptstyle\sim$}}

\newcommand{\equ}[1]{(\ref{#1})} 
\newcommand{\eq}{\begin{equation}} 
\newcommand{\eqn}[1]{\label{#1}\end{equation}} 
\newcommand{\eea}{\end{eqnarray}} 
\newcommand{\eqa}{\begin{eqnarray}} 
\newcommand{\eqan}[1]{\label{#1}\end{eqnarray}} 
\newcommand{\ba}{\begin{array}} 
\newcommand{\ea}{\end{array}} 
\newcommand{\eqac}{\begin{equation}\begin{array}{rcl}} 
\newcommand{\eqacn}[1]{\end{array}\label{#1}\end{equation}} 
 
\newcommand{\ab}{{\rm a}} 
\newcommand{\cm}{{\cal M}}

\newcommand{\hO}{{\hat\Omega}} 
\newcommand{\hA}{{\hat A}} 
\newcommand{\hE}{{\hat E}} 
 
\newcommand{\hd}{{\hat d}}

\newcommand{\hXI}{{\hat\Xi}} 

\newcommand{\hLiexi}{{\LL_\hXI}} 
\newcommand{\dth}{\pa_\th} 
\newcommand{\tp}{{\tilde\p}} 
\newcommand{\tchi}{{\tilde\chi}} 
\newcommand{\tf}{{\tilde\f}} 
\newcommand{\tvf}{{\tilde\vf}} 
 
\newcommand{\ep}{\varepsilon} 
\newcommand{\epp}{{\varepsilon}^{\prime}} 
\newcommand{\xip}{{\xi}^{\prime}} 
\newcommand{\re}{{\rm e}} 
\newcommand{\ds}{{\displaystyle}}

\begin{document} 
\thispagestyle{empty}

\hspace*{\fill}{{\normalsize 
\begin{tabular}{l}
{\sf LYCEN 2004-01}  \\
{\sf UFES-DF-OP2003/6}  \\
%{\sf \today} \\
{\sf May 2004}\\
{\sf Slightly expanded version}\\
\end{tabular}   
 }}

\bigskip 
\bigskip 

\begin{center} 
{\LARGE {\bf Symmetries and Observables}}
\end{center}
\begin{center} 
{\LARGE {\bf  in Topological Gravity}}
%\end{center}
 
\vspace{5mm}

{\large 
Clisthenis P. Constantinidis$^{*}$, 
Aldo Deandrea$^{**}$, 
Fran\c cois Gieres$^{**}$
%\footnote{Supported in part by the 
%{\it Conselho Nacional de
%Desenvolvimento Cient\'\i fico e Tecnol\'ogico (CNPq -- Brazil)}.},
\es Matthieu Lefran\c cois$^{**}$ and Olivier Piguet$^{*}$ 
}
\end{center}
\vspace{5mm}

\noindent $^{*}$ {\it Universidade Federal do Esp\'{\i}rito Santo 
(UFES), 
CCE, Departamento de F\'{\i}sica, Campus Universit\'ario
de Goiabeiras, BR-29060-900 - Vit\'oria - ES (Brasil).}
\vspace{1mm}

\noindent $^{**}$ {\it 
Institut de Physique Nucl\'eaire,
Bat.~Paul Dirac, 
Universit\'e C. Bernard Lyon 1, \\
4, rue Enrico Fermi,
F - 69622 - Villeurbanne Cedex (France).}

\vspace{3mm}

{\tt E-mails: clisthen@cce.ufes.br, 
deandrea@ipnl.in2p3.fr, \\
gieres@ipnl.in2p3.fr, 
lefrancois@ipnl.in2p3.fr,
opiguet@yahoo.com}

%%%%%%%%%%%%%%%%%%%%%%%%%%%%%%%%%%%%%%%%%%%%%%%%%%%%%%%%%%%%%%%
\vspace{2.5cm}

{\small 
\noindent
{\bf Abstract:}
After a brief review of topological gravity,
we present a superspace approach to this theory.
This formulation allows us to recover in a natural 
manner various known results and to gain some insight
into the precise relationship between different
approaches to topological gravity.
Though the main focus of our work 
is on the vielbein formalism, we also
discuss the metric approach and its relationship
with the former formalism. 
}
 
\bigskip
\bigskip
\bigskip

%%%%%%%%%%%%%%%%%%%%%%%%%%%%%%%%%%%%%%%%%% 

\newpage

\tableofcontents

%%%%%%%%%%%%%%%%%%%%%%%%%%%%%%%%%%%%%%%%%%%%%%%%%%%%%%%%
%%%%%%%%%%%%%%%%%%%%%%%%%%%%%%%%%%%%%%%%%%%%%%%%%%%%%%%%
%%%%%%%%%%%%%%%%%%%%%%%%%%%%%%%%%%%%%%%%%%%%%%%%%%%%%%%%
\newpage 

\setcounter{page}{1}

\section{Introduction}

Topological field theories of Witten-type
have been introduced some fifteen
years ago~\cite{ewitten} and have been widely 
studied ever since.
In recent years, they have gained particular attention
in relation with non-topological field theories,
most notably with non-perturbative quantum 
gravity, e.g. see ref.~\cite{smolin}. 
While topological Yang-Mills theories are pretty well
understood by  now~\cite{qft},
this is not true to the same extend for topological gravities
due to the presence of diffeomorphisms. 
The complexity of symmetry algebras and Lagrangians,
as well as  the variety of possible approaches
for topological gravity also 
makes it difficult to compare the 
results obtained using different approaches or formalisms.
Let us 
shortly expand 
on these points.

%as illustrated by recent recent work~\cite{bautan1, spence}.
A topological gravity theory can be constructed 
by gauge fixing an action that is a
topological invariant. 
Alternatively,  it can be introduced by twisting
an extended supergravity 
theory\footnote{In the present work, we have in mind Lagrangian models
as concrete realizations of topological theories.
We do not touch upon the issue of defining cohomological 
theories in the most general way nor do we address 
the question
of whether or not the Lagrangian versions of these 
theories can always be constructed 
by twisting some extended supersymmetric model.}. The latter theories
involve diffeomorphisms and 
local supersymmetry transformations and 
thereby have a sensibly 
more complex structure than super Yang-Mills 
theories.

We now review briefly the different formulations
of topological gravity 
which have been considered in the past 
so as to situate the present work. 
The first papers on topological gravity 
were devoted to the construction of the model
\cite{witten, montano}, while many of the subsequent and 
recent papers \cite{mp}-\cite{mt} 
were rather concerned with the determination of non-trivial 
observables.
Some of the early papers view topological gravity as a 
topological version of 
{\em Weyl (conformal) gravity}
\cite{witten, perry}, 
but these theories do not allow for non-trivial observables.
The remaining work is related to  
ordinary {\em Einstein gravity}. 
The construction of topological gravity 
by a twist of extended supergravity
\cite{anselmi, bautan1, spence} has led to the  study of 
topological {\em Einstein-Maxwell theory}
since extended supergravity theories involve 
a Maxwell field (the so-called graviphoton)
in addition to the vielbein fields. 
Topological gravity can also be viewed as a BF-type model
and has been studied from this point of view 
in a series of papers~\cite{bfgrav}.

Though most works on topological gravity concern space-time manifolds of 
dimension two or four, 
generalizations to higher dimensions have recently been 
introduced~\cite{bautan2}.

Just as for ordinary gravity, different geometric formulations 
can be -- and indeed have been -- developed for topological gravity. 
The most common one is the metric approach: it 
relies only on the metric tensor field and general coordinate 
transformations
as symmetries. If the metric is decomposed with respect to 
vielbein fields, local 
Lorentz transformations also appear as  symmetries (second order formalism).
In addition to the vielbein fields, 
one can consider an independent 
Lorentz connection as basic variable 
(first order formalism) \cite{myers1}.
For ordinary gravity, 
the latter formalism is equivalent to the standard metric approach 
once the connection has been 
eliminated in terms of the vielbein by requiring the torsion to vanish. 
The equivalence also holds for topological gravity, 
but the comparison is more subtle due to the presence 
of extra symmetries\footnote{
We note that, although topological gravity can be formulated 
without vielbein fields, the latter necessarily appear 
-- due to the presence of spinor fields --
in the extended
supergravity theory from which the topological model 
arises by virtue of a twist.}.

Topological field theories of Witten-type involve 
one or several shift symmetries. 
This kind of invariance can be viewed as a relic of 
supersymmetry transformations characterizing 
the extended supersymmetric 
theories 
from which topological models may be constructed
by performing a twist. 
Thus, the shift invariance is also referred to as 
{\em supersymmetry transformation} and it can 
be described conveniently in a superspace 
with an odd coordinate~\cite{horne, osvb} (or several odd 
coordinates for more complex models~\cite{opetal}). 
This formulation, which has been explored previously
for topological Yang-Mills theories, 
allows to obtain the  symmetries, Lagrangian, etc., 
in a compact form 
and to apply the standard
methods of supersymmetry to topological models. 
In particular, standard results on the ordinary BRST cohomology can 
be used~\cite{bcglp} to determine the 
{\em equivariant cohomology}
describing the observables of 
topological field theories~\cite{osvb, kanno} . 
For the case of topological gravity, 
some partial results exist concerning the symmetries
and the Lagrangian in two dimensions~\cite{yuwu}.

The present paper has two parts. 
The first part (and the appendix)
deals with 
previous work on the symmetries and observables
of topological Einstein-Maxwell theory. 
In our presentation, we 
have tried to be geometric and concise,
and to clarify the relationship between different formulations
considered in the literature.  
Apart from the known observables related to the 
topological invariants involving curvature, 
we construct new observables related to a
topological invariant which involves torsion 
and which is not widely known.
In the sequel, we develop a superspace approach 
which leads to 
a complete off-shell formulation for the symmetries.
Simple field redefinitions 
allow us to recover the results discussed in the first part. 
Since our superspace approach explicitly 
involves local supersymmetry transformations
(parametrized by a single odd variable),
it also allows us to compare directly
with the on-shell results which have previously been 
obtained by twisting extended 
supergravity transformations~\cite{anselmi, spence}.
In an appendix, we discuss the metric approach 
and compare with the results obtained 
for the symmetries and observables within the vielbein formalism.
Though the metric approach has 
the advantage of introducing a minimal number of fields, 
it is harder to tackle due to the shift transformations
which act on the metric tensor field and thereby on 
covariant indices. 
We hope that our study will contribute to a better understanding
of the general structure of 
a certain number of 
results and of the 
precise relationship between different approaches to topological
gravity.

%Since our local supersymmetry transformations are parametrized
%by a single odd variable, 
%In our superspace approach, the shift transformations
%of topological gravity appear under the disguise of 
%local supersymmetry transformations.
%One of our aims is to make some progress in this direction
%beyond earlier work on this subject~\cite{myers1}.

%%%%%%%%%%%%%%%%%%%%%%%%%%%%%%%%%%%%%%%%%%%%%%%%%%%%%%%%
%%%%%%%%%%%%%%%%%%%%%%%%%%%%%%%%%%%%%%%%%%%%%%%%%%%%%%%%
%%%%%%%%%%%%%%%%%%%%%%%%%%%%%%%%%%%%%%%%%%%%%%%%%%%%%%%%

\section{Topological gravity} 

After specifying the geometric framework, we will discuss
the symmetries and observables for topological gravity
within the first order formalism.
The reinterpretations to be made in the second order
formalism 
will be commented upon thereafter.

\subsection{Geometric setting}
 
The geometric set-up and the symmetry algebras presented
in the sequel are well defined in any space-time dimension $d$.
We will only specify the dimension for the discussion of observables
 where we focus on the 
%and address more particularly the examples of 
dimensions two and four (subsection \ref{observables}).
Thus, the geometric arena is a real 
$d$-dimensional pseudo-Riemannian manifold ${\cal M}_d$,
the local coordinates being denoted by 
$x=(x^{\mu})_{\mu = 0,...,d-1}$.
Let us briefly recall some geometric notions and results 
that we will use in the sequel \cite{bb, bertlmann, bf}. 

%%%%%%%%%%%%%%%%%%%%%%%%%%%%%%%%%%%%%%%%%%%%%%%%%%%%%%%%
\paragraph{BRST formalism}
Within the BRST formalism, the parameters of infinitesimal 
symmetry transformations are turned into ghost fields.
The latter have ghost-number $g=1$ while the basic fields
appearing in the invariant action have
a vanishing ghost-number.
 The Grassmann parity of an object is given by the 
parity of its {\it total degree} defined as the sum $p+g$ of 
its form degree $p$ and  ghost-number $g$.
All commutators and  brackets are assumed to be graded 
according to this grading.  

The BRST operator, which is denoted by $\cas$, acts 
on the algebra of fields as an antiderivation
which increases the ghost-number (and thus the total degree) 
by one unit. It is assumed to anticommute with the 
exterior derivative $d$.

%%%%%%%%%%%%%%%%%%%%%%%%%%%%%%%%%%%%%%%%%%%%%%%%%%%%%%%%
\paragraph{Vector fields, inner product and Lie derivative}  
For a vector field $w=w^\m \pa_\m$  
  on $\MM_d$,  
the {\em total degree} is given by its ghost-number which 
we denote by $[w]$. 
It is said to be {\em even (odd)} if $[w]$ is even (odd).  
 
The Lie bracket $[u,v]$ of two vector fields $u$ and $v$   
is again a vector field: this bracket is assumed to be  
graded so that its  
components are  given by  
\eq
[u,v]^\m = u^\n\pa_\n v^\m  -(-1)^{[u][v]}   v^\n\pa_\n u^\m  
\ .  
\eqn{grlbr}
 
    The  {\em interior product}  
 $i_w$ with respect to the vector field 
  $w= w^\m \pa_\m$ is defined in local coordinates by 
$i_w\vf = 0$
for $0$-forms and $i_w (dx^\m ) = w^\m$.  
  If $w$ is even, the operator  $i_w$  acts as an antiderivation  
(odd operator), otherwise it acts  
as a derivation (even operator).

The {\em Lie derivative} $\LL_w$ 
with respect to $w$  
acts on differential forms according to  
\eq 
\LL_w \equiv [i_w,d \, ] = i_w d  +(-1)^{[w]}  d i_w  
\eqn{Lie-deriv} 
and we have the graded commutation relations 
\begin{equation} 
 \lc \LL_u,\LL_v \rc = \LL_{[u,v]} 
\quad  ,\quad  
\lc \LL_u , i_v \rc  =  i_{[u,v]}
 \, . 
\label{li}
\end{equation}

In the following, the quantity $\xi = \xi^{\mu} \pa_{\mu}$ 
always denotes a vector
field of ghost-number $1$ (representing the ghost for
diffeomorphisms). We then have  
the following identities involving the vector fields 
$\xi$ and  
$\xi^2 \equiv {1 \over 2} [ \xi , \xi ]$ as well as  
the previously introduced
operators
(in particular the exponential $\re ^{i_{\xi}}$of the 
linear operator $i_{\xi}$): 
\eq\ba{l}
\re ^{i_{\xi}} ( X \, Y )  =  \lp  \re ^{i_{\xi}} X\rp\lp  \re
^{i_{\xi}}Y\rp
\es 
{\rm e}^{- i_{\xi}} d \re ^{i_{\xi}}  =  d - {\cal L}_{\xi} - i_{\xi^2}
\es
\left[ \cas, \re ^{i_{\xi}} \right]  
=  i_{\cas \xi}\, \re ^{i_{\xi}} 
\quad  , \quad \left[ \cas, \re ^{- i_{\xi}} \right]  
= - i_{\cas \xi}\, \re ^{- i_{\xi}} 
\ .
\ea\eqn{ident1}

%%%%%%%%%%%%%%%%%%%%%%%%%%%%%%%%%%%%%%%%%%%%%%%%%%%%%%%%%%%%

\subsection{First order formalism} 

In the first order formalism of the theory, 
the basic variables are the vielbein $1$-forms 
$e = (e^a)_{a=0,...,d-1}$ and the Lorentz connection $1$-form 
$\omega = (\omega ^a_{\ b})$. 
The tangent space indices $a,b,...$ are raised or lowered using 
the constant tangent space metric $(\eta_{ab})$ which can be of 
Minkowskian or of Euclidean signature. 
In the following, we will use the matrix notation $e, \omega, \dots$
so as to avoid spelling out 
the tangent space indices $a,b,\dots$ 

Since topological gravity  is expected to 
originate from a  
twisted extended supergravity theory, we also introduce
an Abelian (Maxwell or $U(1)$) gauge connection $1$-form $\ab$,
the so-called {\em graviphoton} field that generally 
appears in 
extended supergravity theories.

%As we shall see howevever at the end
%of section \ref{Superspace-approach}, the same ``graviphoton'' may
%originate in the superspace formalism as a component of the super-vielbein.

The respective field strengths of $e, \, \omega $
and $\ab$ are the torsion $2$-form  
$T = De \equiv de +\omega  e$, the curvature $2$-form 
$R = d\omega + {1 \over 2} [ \omega  , \omega  ]$
and the Abelian curvature $2$-form 
$F_{\ab}= d\ab$.
They satisfy the Bianchi identities
\begin{eqnarray*}
 D R \!\!\!&=&\!\!\! 0
\, ,  \ \qquad {\rm where} \ \; DR \equiv dR + [ \omega ,R]
\\
 D T \!\!\!&=&\!\!\! Re
\, , \ \; \quad {\rm where} \ \; DT \equiv dT + \omega T 
\\ 
dF_{\ab} \!\!\!&=&\!\!\! 0 \, . 
\end{eqnarray*}

The basic symmetries are diffeomorphisms and 
local Lorentz transformations parametrized in a BRST setting by 
ghosts 
$\xi = \xi^{\mu} \partial _{\mu}$ and $c = (c^a_{\ b})$,
as well as local $U(1)$ transformations
parametrized by a ghost $u$.
Note that both 
$\omega$ and $c$ take their values in the Lie algebra of the 
Lorentz group, i.e.
$\omega _{ab}=- \omega _{ba}$ and $c_{ab} = - c_{ba}$.

\subsubsection{Horizontality conditions}\label{hhh} 
   
   We introduce the {\em generalized differential}
$\hat{d} = d+\cas$ and the {\em generalized fields} 
\cite{exp-xi, bertlmann}
\begin{equation}
\label{gen}
\begin{array}{lll} 
\hat{\omega } \equiv   {\rm e}^{i_{\xi}} (\omega +c)  
= \omega  + c + i_{\xi} \omega \, ,
& \quad &  
\hat{e} \equiv  {\rm e}^{i_{\xi}} e
= e +  i_{\xi} e 
\\
\hat{\ab} \equiv  {\rm e}^{i_{\xi}} (\ab +u)  
= \ab + u + i_{\xi} \ab \, , 
& \quad &  
\\
& \quad &  
\\
\hat{R} \equiv  \hat{d} \hat{\omega } + {1 \over 2} 
[ \hat{\omega }, \hat{\omega } ]  \, ,
& \quad &  
\hat{T} \equiv  \hat {D} \hat{e} 
\, = \, 
\hat{d} \hat{e} + \hat{\omega } \hat{e} 
\\
\hat{F}_{\ab} \equiv \hat{d} \hat{\ab}
\, , 
& \quad &  
\end{array}
\end{equation}
which imply Bianchi identities for the generalized curvature and 
torsion 
forms:  
\begin{equation}
\label{genbi}
 \begin{array}{lll} 
\hat D \hat R = 0 \, , 
& \quad &  
\hat  D  \hat T =  \hat R  \hat e
\\
 \hat d \hat F_{\ab} = 0 \, . 
& \quad &  
\end{array}
\end{equation}
By expanding the generalized $2$-forms 
$\hat R, \hat T$ and $\hat F_{\ab}$ 
with respect to the ghost-number
we find 
\begin{equation}
\label{expcurv}
\hat R = R_2^0 + R_1^1 + R_0^2 \, , 
\quad {\rm with} \ \ \left\{
\begin{array}{l}
R_2^0 = R 
\\
R_1^1 = \cas \omega + D c_{\xi} 
\qquad (\,  c_{\xi} \equiv c + i_{\xi} \omega \, ) 
\\
R_0^2 = \cas c_{\xi} + c^2_{\xi} 
\, ,
\end{array} 
\right.
\end{equation} 
as well as similar expressions for $ \hat T$ and $\hat F_{\ab}$.

The BRST transformations of all space-time 
fields then follow from  
relations (\ref{genbi}) by imposing a horizontality condition,
i.e. by specifying $R_1^1$ and $R_0^2$ 
($R_2^0$ being necessarily equal 
to the curvature $2$-form $R$)
and by specifying the corresponding components 
of $\hat T$ and $\hat F_{\ab}$. 
For topological gravity, 
one imposes the 
following {\em horizontality conditions} \cite{bautan1}  
which generalize those of topological Yang-Mills 
theories\footnote{The geometrical interpretation 
of horizontality conditions
for ordinary and topological Yang-Mills theories
are discussed in references \cite{bertlmann} 
and \cite{bbrt}, respectively.}:
\begin{equation}
\label{hc}
 \begin{array}{lll} 
\hat{R} =  {\rm e} ^{i_{\xi}} ( R + \tp + \tilde{\phi} ) \, , 
& \quad &  
\hat{T} =  {\rm e} ^{i_{\xi}} (T + \psi + \phi) 
\\
\hat{F}_{\ab} =  {\rm e} ^{i_{\xi}} 
( F_{\ab} + \eta + t ) \,  . 
& \quad &  
\end{array}
\end{equation}
Here, $\tp^a_{\ b}, \, \psi^a$ 
and $\eta$
are $1$-forms with ghost-number $1$,  
while  $\tilde{\phi}^a_{\ b}, \, \phi ^a$ 
and $t$
are $0$-forms with ghost-number $2$. 
The fields $\tp$ and $\tilde{\phi}$ are Lorentz algebra-valued, i.e. 
$\tp_{ab} = - \tp_{ba}$ and 
$\tilde{\phi}_{ab} = - \tilde{\phi}_{ba}$.

By substituting the expansion (\ref{expcurv})
and the analogous expansions for $\hat T$ and $\hat F_{\ab}$
into (\ref{hc}), we see that the ghost-number $2$ fields 
 $\tilde{\phi}^a_{\ b}, \, \phi ^a$ 
and $t$ appear in the $\cas$-variations of the ghost fields
so that they represent {\em ghosts for ghosts}. 
Their appearance expresses the reducibility of the 
resulting symmetry algebra,
see remark (i) below.

 The action of the operator $\re ^{i_{\xi} }$ can be factorized 
in all terms of 
equations (\ref{genbi}) and (\ref{hc}) by virtue of the following 
operatorial relation 
\cite{bb, bf}
which results from  equations (\ref{ident1}):
\begin{equation}
\label{oprel}
(d+\cas) {\rm e}^{i_{\xi}} = 
{\rm e}^{i_{\xi}}( d+ \cas  - 
\LL_{\xi} + i_{\cas\xi - \xi^2})
\, .
\end{equation}
Let $\varphi \equiv \varphi^{\mu} \pa_{\mu}$ denote 
the vector field $\cas\xi - \xi^2$ 
which appears on the right-hand-side 
and which is 
of ghost-number $2$. 
The requirement of 
nilpotency for the variation $\cas\xi$ then implies that the 
$\cas$-variation
of the vector field 
$\varphi$ is given by its Lie derivative:
%$ \LL_\xi \varphi = [\xi, \varphi]$, i.e. 
\begin{equation}
\label{brs2}
\cas\xi = \xi^2 + \varphi 
\quad , \quad 
\cas \varphi = [\xi, \varphi]
\, .
\end{equation}   
Since $\varphi$ describes a local shift of the 
diffeomorphism ghost, it parametrizes 
{\em vector supersymmetry} transformations~\cite{op}.
By expanding equations (\ref{hc}) and (\ref{genbi}) 
with respect to the ghost-number, we get  
the $\cas$-variations of all fields as well as 
the relation $\phi = i_{\varphi} e$. 
The latter is equivalent to the relation 
$\varphi^{\mu} = \phi^a e_a^{\mu}$ 
if we assume the vielbein to be invertible.
Thus, it expresses the variable 
$\varphi$ in terms of $\phi$ and the inverse vielbein
$e_a ^{\mu}$.
Since it is the field $\vf$, rather than $\f$, that
appears explicitly  in the BRST transformation of the
diffeomorphism ghost $\xi$, 
it is actually necessary to assume 
the vielbein to be invertible at this stage.

After the change of variables
\begin{eqnarray}
\label{cvar}
\tilde{\phi}  \!\!\!&&\!\! \to \ \tvf := \tilde{\phi} -  i_{\varphi} \omega 
\\
t  \!\!\!&& \!\! \to \ \tau := t - i_{\varphi} \ab
\, , 
\nonumber
\end{eqnarray}   
the {\em $\cas$-variations} 
of the basic fields 
take the simple form 
\begin{equation}
\label{brs1}
\begin{array}{lll}
\cas e = \LL_\xi e  - ce + \p 
\ , & \quad & 
\cas \xi =  \xi^2 + \varphi 
\\
\cas \p =  \LL_\xi \p - c\p 
- \LL_\vf e + \tvf e 
\ , & \quad & 
\cas  \varphi = [\xi, \varphi]
\\
$\quad$ && 
\\
\cas \om =  \LL_\xi \om  -Dc  +  \tp 
\ , & \quad & 
\cas c =  \LL_\xi c - c^2 +  \tvf
\\
\cas \tp =  \LL_\xi \tp - [c,\tp] 
- \LL_\vf \omega -D\tvf 
\ , & \quad & 
\cas \tvf =  \LL_\xi \tvf - [c,\tvf] -\LL_\vf c 
\\
$\quad$ && 
\\
\cas \ab =  \LL_\xi \ab  -du  +  \eta 
\ , & \quad & 
\cas u = \LL_\xi u + \tau 
\\
\cas \eta =  \LL_\xi \eta
- \LL_\vf \ab - d\tau 
\ , & \quad & 
\cas \tau  =  \LL_\xi \tau  -\LL_\vf u 
\end{array}
\end{equation}
and those of the field strengths read as  
\begin{equation}
\label{sto}
\begin{array}{lll}
\cas R =  \LL_\xi R - [c,R]  -D\tp
\ , & \quad & 
\cas T =  \LL_\xi T - cT + \tp e -D\p  
\\
\cas F_{\ab} =  \LL_\xi F_{\ab}   -d \eta
\, .
\end{array}
\end{equation}
By construction, the so-defined $\cas$-operator is nilpotent
and the results coincide with those 
given in 
references~\cite{bautan1, mt}, 
except for some terms in the $\cas$-variations of $\tvf$
and $T$ which are missing in \cite{bautan1} and \cite{mt},
respectively, and which ensure the nilpotency of $\cas$.
%the term $\LL_{\vf}c$ in the $\cas$-variation
%of $\tvf$ which was obviously forgotten in~\cite{bautan1} 
%and which ensures the nilpotency of $\cas$.

\paragraph{Remarks:}

\noindent 
{\bf (i)}
It is easy to understand the origin of all terms appearing 
in the transformation 
laws (\ref{brs1}). 
We only consider the gravitational sector since the argumentation
for the Maxwell sector proceeds along the same lines.
The $\cas$-variations of the 
basic fields $e$ and $\omega$ describe diffeomorphisms
(parametrized by $\xi$), 
local Lorentz transformations (parametrized by $c$)
and the {\em topological} (or {\em shift})
{\em symmetry} (parametrized by $\psi$ and $\tp$),
which is characteristic for  topological field
theories of Witten-type. 
Obviously, the  $\cas$-variation of $e$ is reducible: 
$\cas e$ is invariant under a shift $\delta \xi = \varphi$ 
which comes together with the transformation 
$\delta \psi = - \LL _{\varphi} e$. Furthermore, 
it is invariant under the  shift $\delta c= \tilde{\varphi}$ 
that goes together with $\delta \psi = \tvf e$. 
Similarly, the  $\cas$-variation of $\omega$ is 
 invariant under the shifts
$\delta \xi = \varphi$  and   $\delta c= \tilde{\varphi}$ 
accompanied by the  transformations
 $\delta \tp  = - \LL _{\varphi} \omega$ and 
$\delta \tp = D \tvf$, respectively. 
The BRST algebra can then be completed by assuming all fields 
to change linearly under 
Lorentz transformations and to transform 
with the Lie derivative
under diffeomorphisms: this leads to the 
reducibility of $\cas c$ under the shift $\delta \xi = \varphi$ 
accompanied by the transformation 
$\delta \tvf = - \LL _{\varphi} c$. 
Thus, all terms have a natural interpretation
and the signs are simply a  matter of nilpotency. 

\smallskip

\noindent 
{\bf (ii)}
The 
{\em supersymmetry} or {\em shift operator} is given by 
\[
\tilde Q \equiv \cas - \LL_{\xi} - \delta_c ^{(L)} 
- \delta_u ^{(M)} 
\]
where $\delta_c ^{(L)} $ 
and $\delta_u ^{(M)}  $
denote, respectively, infinitesimal Lorentz
and Maxwell transformations.
When applied to the basic fields and ghosts,  it satisfies
\[
\tilde Q ^2 = - \LL_{\varphi}  - \delta_{\tvf} ^{(L)} 
- \delta_{\tau} ^{(M)} 
\ ,
\]
i.e. $\tilde Q$ is nilpotent up to infinitesimal diffeomorphism,
Lorentz and Maxwell transformations with parameters $\varphi, \, \tvf$
and $\tau$, respectively.

\smallskip

\noindent 
{\bf (iii)}   
An arbitrary shift $\psi_{\mu}^a$ of the vielbein $e_{\mu}^a$ does not
preserve the positivity of the determinant of the metric
\cite{mp}. 
This ``problem'' can be solved \cite{mp}
by assuming that the shift of the vielbein
is described by a local infinitesimal gauge transformation
associated to the general linear group
$GL(n,{\bf R})$, i.e. by assuming $\psi_{\mu}^a$
to be of the form 
$G^a _{\, b} e^b_{\mu}$ with 
$( G^a _{\, b}) \in GL (n,{\bf R}) $.
The BRST algebra then takes
a form which is quite similar to (\ref{brs1}).

However, topological invariants
are inert under {\em arbitrary} shifts of the metric (or vielbein)
and therefore the positivity of the determinant of the 
metric does not necessarily have to be  imposed 
at this point.

\smallskip

\noindent 
{\bf (iv)}   
By changing generators 
according to reparametrizations of the form 
$c ^{\prime} =c + i_{\xi} \omega, \,  
\psi ^{\prime} = \psi + (i_{\xi} \omega ) e, \dots$, 
the BRST algebra (\ref{brs1}) can be cast into 
equivalent forms.  
%Such a form appears to underly the algebra 
%presented in reference \cite{nso}, though the given $\cas$-variations
%fail to be nilpotent. 
One such reformulation can be obtained from a different reading
of the generalized fields and horizontality conditions.
This parametrization naturally appears in the group manifold 
approach and the associated  rheonomic
parametrization of curvatures \cite{anselmi}.
In fact, let us read the generalized fields (\ref{gen}) as 
\begin{eqnarray} 
\hat{\omega } \!\!\!& = &\!\!\!  \omega  + c_{\xi}  \ , 
\qquad {\rm with} \ \, c_{\xi} \equiv c + i_{\xi} \omega 
\nonumber 
\\
\hat{e} \!\!\!& = &\!\!\! 
e + \varepsilon_{\xi}  \ \, , 
\qquad {\rm with} \ \, \varepsilon_{\xi} ^a \equiv \xi ^a
= \xi^{\mu} e_{\mu}^a 
\nonumber 
\\
\hat{\ab} \!\!\!& = &\!\!\! 
\ab + u_{\xi}  \ , 
\qquad {\rm with} \ \; u_{\xi} \equiv u + i_{\xi} \ab 
\ , 
\label{gen6}
\end{eqnarray} 
and the horizontality conditions (\ref{hc}) as 
\begin{equation}
\label{hc6}
\begin{array}{lll} 
\hat{R} =  R + \tp_{\xi} + \tilde{\phi}_{\xi}  \, , 
& \quad & 
\hat{T} = T + \psi_{\xi} + \phi_{\xi} 
\\
\hat{F}_{\ab} =
F_{\ab} + \eta_{\xi} + t_{\xi} 
\,  ,
& \quad &  
\end{array}
\end{equation}
with $\tp_{\xi} = \tp + i_{\xi} R, \, 
\tilde{\phi}_{\xi}  = \tilde{\phi} + i_{\xi}\tp +{1 \over 2} 
 i_{\xi}  i_{\xi} R$, etc. 
Expansion of relations (\ref{hc6})
with respect to the ghost-number immediately
yields the $\cas$-variations in their 
``Lorentz- and Maxwell-covariantized form'':
\begin{equation}
\label{brst6}
\begin{array}{lll}
\cas e =  - D\varepsilon_{\xi}  - c_{\xi} e + \p _{\xi} 
\ , & \quad & 
\cas  \varepsilon_{\xi} =  -c_{\xi} \varepsilon_{\xi} 
+ \phi _{\xi}
\\
\cas \p_{\xi} =  -D \phi _{\xi} - c_{\xi}
\p _{\xi} +  \tp _{\xi} \varepsilon_{\xi} + \tilde{\phi} _{\xi} e
\ , & \quad & 
\cas \phi_{\xi} = - c_{\xi} \phi_{\xi} 
+ \tilde{\phi} _{\xi} \varepsilon_{\xi}
\\
$\quad$ && 
\\
\cas \om =  -Dc_{\xi}  +  \tp _{\xi} 
\ , & \quad & 
\cas c_{\xi} =  - c_{\xi}^2 +  \tilde{\phi} _{\xi}
\\
\cas \tp _{\xi} =  -D\tilde{\phi} _{\xi} 
- [c_{\xi},\tp _{\xi} ] 
\ , & \quad & 
\cas \tilde{\phi} _{\xi} =  - [c_{\xi},\tilde{\phi} _{\xi}]  
\\
$\quad$ && 
\\
\cas \ab =  -du_{\xi}  +  \eta_{\xi} 
\ , & \quad & 
\cas u_{\xi} = t_{\xi} 
\\
\cas \eta_{\xi} =  -d t_{\xi} 
\ , & \quad & 
\cas t_{\xi}  =  0
\end{array}
\end{equation}
and 
\begin{equation}
\label{sto6}
\begin{array}{lll}
\cas R =  -D\tilde{\psi} _{\xi} - [c_{\xi},R]  
\ , & \quad & 
\cas T =  -D \psi _{\xi} - c_{\xi}T +R \varepsilon_{\xi}
+ \tp _{\xi} e 
\\
\cas F_{\ab} =  -d \eta_{\xi}
\ .
\end{array}
\end{equation}
These expressions coincide with those of 
reference~\cite{anselmi}. 
An advantage of this para\-metrization consists of the fact that 
the field $\tilde{\phi} _{\xi} $ simply transforms like a 
commutator, exactly as the ghost for ghost in topological Yang-Mills
theories. Henceforth, BRST invariant polynomials in this variable 
are generated by $\tr (\tilde{\phi} _{\xi} )^n $
with $n=1,2, \dots$
We will come back to this point in the next subsection.

In conclusion, we mention one more change of generators which
allows to cast the BRST algebra into another equivalent form
which  appears more or less explicitly in the early works on the
subject,  e.g. see references~\cite{nso, oda, myers1}.
In fact, by virtue of the reparametrization
$c \to c_{\xi} = c +i_{\xi} \omega$ and 
$\tilde{\varphi} \to \tilde{\phi} = \tilde{\varphi} + i_{\varphi} \omega$,
the $\cas$-variations of the gravitational sector, 
as given by equations (\ref{brs1}), take the following form 
involving the Lorentz-covariant Lie derivative
$L_{\xi} \equiv i_{\xi} D - D i_{\xi}$:
\begin{equation}
\label{brs4}
\begin{array}{lll}
\cas e = L_\xi e  - c_\xi e + \p 
\ , & \quad & 
\cas \xi =  \xi^2 + \varphi 
\\
\cas \p =  L_\xi \p - c_\xi \p 
- L_\vf e + \tf e 
\ , & \quad & 
\cas  \varphi = [\xi, \varphi]
\\
$\quad$ && 
\\
\cas \om =  i_\xi R  -Dc_\xi   +  \tp 
\ , & \quad & 
\cas c_\xi  =  i_\xi \tp +{1 \over 2} \, i_\xi i_\xi R 
- c_\xi ^2 +  \tf
\\
\cas \tp =  L_\xi \tp - [c_\xi,\tp] 
- i_{\vf} R -D\tf 
\ , & \quad & 
\cas \tf \, = \, L_\xi \tf \, - \, [c_\xi,\tf] \, - \, i_\vf \tp 
\, . 
\end{array}
\end{equation}

%%%%%%%%%%%%%%%%%%%%%%%%%%%%%%%%%%%%%%%%%%%%%%%%%%%%%%%%%%

\subsubsection{Observables}\label{observables}

The construction of observables for topological gravity 
is based on gauge invariant polynomials 
of the curvature 
form (e.g. $\tr \{ R^k \}$ with $k =1,2,\dots$)
and of the torsion form.
A topological invariant involving torsion has first 
been introduced 
in four dimensions by Nieh and Yan~\cite{ny} and has been 
further discussed by the authors of reference~\cite{zanelli} 
who also constructed higher dimensional 
generalizations.
The following discussion of the cases $k=2$ and $k=1$ 
applies to manifolds which are at least of {\em dimension four}
and {\em two}, respectively.  

\paragraph{Case $k=2$ :}

We  first consider the gravitational sector and 
comment on the Maxwell sector thereafter.

The {\em Pontryagin density}, i.e. 
the $4$-form 
$W ^0 _4 \equiv -\ds{1 \over 2} \, \tr \{ RR\}$ is closed by 
virtue of the Bianchi identity $DR=0$:
\[
d  W^0 _4 = - \tr \{ (DR)R \} = 0 
\, .
\]
Accordingly, the generalized 
$4$-form 
$\hat{W}  \equiv -\ds{1 \over 2} \, \tr \{ \hat R \hat R \}$ 
is annihilated by the 
generalized differential $\hat d \equiv d +\cas$, i.e. 
\begin{equation}
\label{sdes}
\cas \hat{W} = - d\hat{W}
\, .
\end{equation}
By substituting (\ref{hc}) into $\hat{W}$ and 
expanding with respect to the ghost-number, 
we obtain 
 \begin{equation}
\label{exgn}
\hat{W} = -\ds{1 \over 2} \,   \re^{i_{\xi}} 
\, \tr \{ ( R + \tp + \tilde{\phi} )  ( R + \tp + \tilde{\phi} ) \} 
 \, = \, 
 \sum _{k=0} ^4 \, W ^k _{4-k} (\xi)
\, .
\end{equation} 
Here, the $\xi$-dependence is explicitly given by 
 \begin{equation}
\label{poly}
W^k _{4-k} (\xi) = 
 \sum _{n=0} ^k \, \ds{1 \over n!} \, (i_{\xi})^n  \, W^{k-n} _{4-k+n}
\, , 
\end{equation} 
where 
the polynomials $W^{\kappa}_{4- \kappa} $ 
appearing on the right-hand-side 
have the same form 
as the {\em Donaldson  polynomials} in topological 
Yang-Mills theory:
\begin{eqnarray}
W^0 _4\!\!\!&=&\!\!\!
-\ds{1 \over 2} \, \tr \{ RR\} 
\quad , \quad 
W^1 _3 = -  \, \tr \{ \tp R\} 
\quad , \quad 
W^2 _2 =
-\, \tr \{\tilde{\phi}R + \ds{1 \over 2} \tp \tp \} 
\nonumber 
\\
W^3 _1
\!\!\!&=&\!\!\!
-\, \tr \{\tilde{\phi} \tp \} 
\quad , \quad 
W^4 _0 = -\ds{1 \over 2} \, \tr \{\tilde{\phi} \tilde{\phi} \} 
\, .
\label{oym}
\end{eqnarray}
In particular, $W^0 _4 (\xi) = W^0 _4 = -\ds{1 \over 2} \, \tr \{ RR\} $.
If one uses the relation 
$\tilde{\phi} = \tilde{\varphi} + i_{\varphi} \omega$
to express $\tilde{\phi}$ in terms of the variable $\tilde{\varphi}$
which appears in the BRST transformations (\ref{brs1}), 
then the polynomials $W_2^2(\xi)$, $W_1^3(\xi)$ and 
$W_0^4(\xi)$ also depend on the shift $\varphi$ of $\xi$. 

By virtue of the relation (\ref{sdes}), the polynomials (\ref{poly}) 
satisfy the descent equations\footnote{We note that 
the $\cas$-variation of $\tilde{\phi}$ is 
given by 
$\cas \tilde{\phi} = \LL_{\xi} \tilde{\phi}
- [c, \tilde{\phi} ] - 
i_{\varphi} \tp $.}
\begin{eqnarray}
 dW^0 _{4} (\xi) \!\!\!&=&\!\!\! 0
\nonumber 
\\
\cas W^k_{4-k} (\xi) + d W^{k+1}_{3-k} (\xi) \!\!\!&=&\!\!\! 0
\qquad {\rm with} \ \, 0 \leq k \leq 3
\nonumber 
\\
\cas W^4 _{0} (\xi) \!\!\!&=&\!\!\! 0
\, .
\label{des4d}
\end{eqnarray}
The polynomials $ W ^k _{4-k} (\xi)$ represent elements of the 
so-called {\em equivariant cohomology}~\cite{osvb, kanno}
of topological gravity.
By contrast to the case of topological Yang-Mills theories, a
ghost associated to gauge transformations, namely the 
diffeomorphism ghost $\xi$, appears in the 
cohomology classes~\cite{myers2}.
However, this ghost does not play the same r\^ole as 
the ghosts $c$ and $u$ associated to Lorentz and Maxwell 
gauge transformations 
(since its action amounts to moving points on the 
space-time manifold)
and its presence is actually 
necessary~\cite{wu}.

In four dimensions, another set of observables can be obtained 
from the {\em Euler class} 
$V^0 _{4} \equiv - {1 \over 2}  
\tr \{ \varepsilon_{abcd} R^{ab}R^{cd} \}$ 
whose integration yields the Euler characteristic. 
One follows the same procedure, i.e. one expands 
$\hat V \equiv  - {1 \over 2}  
\tr \{ \varepsilon_{abcd} \hat R ^{ab} \hat R^{cd} \}$
with respect to the ghost-number as in equation (\ref{exgn}):
$\hat{V} = \sum _{k=0} ^4 \, V ^k _{4-k} (\xi)$. 

Since $\hat d \hat W =0 = \hat d \hat V$, we also have 
$\hat d (\hat W ^m \hat V ^n)=0$ for 
$m,n \in \{0,1,\dots \}$.
By expanding $\hat W ^m \hat V ^n$ with respect to the ghost-number,
one obtains further representatives of the 
cohomology algebra~\cite{thuillier}:
\[
\hat W ^m \hat V ^n
= w_0^{4(m+n)} (\xi) + w_1^{4(m+n)-1} (\xi) + \cdots 
+ w_4^{4(m+n)-4} (\xi)
\, ,
\]
with
\begin{eqnarray}
w_0^{4(m+n)} (\xi) \!\!\! &=&  \!\!\!
\left[ W_0^{4} (\xi) \right]^m \left[ V_0^{4} (\xi) \right]^n
\\
w_1^{4(m+n)-1} (\xi) \!\!\! &=&  \!\!\!
n \left[ W_0^{4} (\xi) \right]^m \left[ V_0^{4} (\xi) \right]^{n-1} 
V_1^{3} (\xi)
+ m  \left[ W_0^{4} (\xi) \right]^{m-1} 
W_1^{3} (\xi)
\left[ V_0^{4} (\xi) \right]^{n} , \, {\rm etc.}
\nonumber 
\end{eqnarray}

An obvious question is whether or not there exist further 
elements in the gravitational sector of the 
equivariant cohomology
which are related to the curvature form.
To address this 
problem, it is useful to recall the 
variables $\tp_{\xi}$ and 
$\tilde{\phi}_{\xi}$ introduced in equation (\ref{hc6})
and to invoke a simple argument put forward in reference~\cite{anselmi}. 
Due to the very definition of $\tp_{\xi}$ and $\tilde{\phi}_{\xi}$,
we can read expression (\ref{exgn}) as
\begin{equation}
\hat{W} = -\ds{1 \over 2} 
\, \tr \{ ( R + \tp_{\xi} + \tilde{\phi}_{\xi} )  
( R + \tp_{\xi} + \tilde{\phi}_{\xi} ) \} 
\, ,
\end{equation}
so that the polynomials $W_{4-k}^k (\xi)$ are 
the Donaldson polynomials (\ref{oym}) with 
$\tp$ and $\tilde{\phi}$ replaced by $\tp_{\xi} $ and 
$\tilde{\phi}_{\xi}$, respectively. 
In particular, we have
$W^4 _{0} (\xi) = 
 -\ds{1 \over 2} \, 
 \tr \{\tilde{\phi} _{\xi} \tilde{\phi} _{\xi}  \} $.
This result is consistent with the comment (made 
after equation (\ref{sto6})) that 
the $\cas$-invariant polynomials in $\tilde{\phi} _{\xi}$
are generated by the Lorentz invariant polynomials
$\tr (\tilde{\phi} _{\xi}) ^n$ with $n=1,2,\dots$
More specifically, for the 
four dimensional case that we consider here, 
the $4 \times 4$ matrix $\tilde{\phi} _{\xi}$
has the four invariants 
$\tr \tilde{\phi} _{\xi}$, $\tr (\tilde{\phi} _{\xi})^2$, 
$\tr (\tilde{\phi} _{\xi})^3$, $\tr (\tilde{\phi} _{\xi})^4$,
but $\tr \tilde{\phi} _{\xi}$ and $\tr (\tilde{\phi} _{\xi})^3$
vanish due to the antisymmetry of the matrix
$\tilde{\phi} _{\xi}$.
Thus, the only non-trivial invariants are $\tr (\tilde{\phi} _{\xi})^2$
and $\tr (\tilde{\phi} _{\xi})^4$, where the latter is generated by 
the former and 
$\tr \{\varepsilon_{abcd} \tilde{\phi}^{ab} _{\xi} 
\tilde{\phi} _{\xi}^{cd}  \}$ due to the identity
\[
\tr (\tilde{\phi} _{\xi})^4 = \ds{1 \over 16}
\left( \tr \{\varepsilon_{abcd} \tilde{\phi}^{ab} _{\xi} 
\tilde{\phi} _{\xi}^{cd}  \} \right)^2 + \ds{1 \over 2} 
\left( \tr (\tilde{\phi} _{\xi})^2 \right)^2
\, .
\]
Thus, for $k=2$, the polynomials 
 $W_0^4 (\xi) \propto  
\tr \{\tilde{\phi} _{\xi} \tilde{\phi} _{\xi}  \}$ 
and $V_0^4 (\xi)  \propto 
\tr \{\varepsilon_{abcd} \tilde{\phi}^{ab} _{\xi} 
\tilde{\phi} _{\xi}^{cd}  \}$
generate the whole cohomology of the 
Lorentz connection sector.
To summarize, the fact that the part of the BRST algebra 
which involves the Lorentz 
connection can be cast into a form that is isomorphic
to the BRST algebra 
of topological Yang-Mills theory allows us to 
invoke the known results concerning the equivariant 
cohomology of the latter, e.g. see ref.~\cite{bcglp}.

Let us now turn to the vielbein part of the gravitational BRST algebra.
Quite remarkably, there exists a local expression 
given by the vielbein and connection fields   
whose integral only depends on the topology,  
i.e. on the global properties of the space-time manifold.
In four dimensions, this topological invariant related to the torsion
is the integral over the Nieh-Yan form~\cite{ny, zanelli}, 
i.e. the $4$-form
\begin{equation}
\label{nieh}
Z_4^0  \equiv - \ds{1 \over 2} (TT - e R e )
=
-\ds{1 \over 2} (T^a T_a - e^a R_{ab} e^b )
\, .
\nonumber
\end{equation}
The latter vanishes if the torsion vanishes (due to the Bianchi identity
$Re =DT$). Furthermore, this form is closed and  locally exact:
$d(eT) = D(eT) = -2 Z_4^0$. 
Accordingly, we can proceed as for the Pontryagin or Euler density,
i.e. exploit the fact that 
the generalized $4$-form 
$\hat Z \equiv - \ds{1 \over 2} (\hat T \hat T - \hat e \hat R \hat e )$
is annihilated by the generalized differential $\hat d = d + \cas$.
Thus, we expand $\hat Z$ with respect to the ghost-number
analogously to the expansion of $\hat W$ in equations (\ref{exgn}) and 
(\ref{poly}). By construction, the polynomials $Z^k_{4-k} (\xi)$ 
appearing in this expansion satisfy 
the descent equations (\ref{des4d}).
Explicit expressions can readily be obtained from those
of $\hat e, \hat T, \hat R$ given in equations (\ref{gen})
and (\ref{hc}).
The results take a concise form when written in terms of the 
reparametrized ghosts 
$\varepsilon_{\xi}, 
\tp_{\xi}, 
\tilde{\phi}_{\xi},
\psi_{\xi}, 
\phi_{\xi}$ 
introduced in (\ref{gen6}) and (\ref{hc6}):
\begin{eqnarray}
Z_4^0 (\xi) \!\!\! &=& \!\!\! -\ds{1 \over 2} (TT - e R e )
\nonumber
\\
Z_3^1 (\xi) \!\!\! &=& \!\!\! - (T \psi_{\xi} 
- \varepsilon_{\xi} R e
-\ds{1 \over 2} e \tp_{\xi} e )
\nonumber
\\
Z_2^2 (\xi) \!\!\! &=& \!\!\! -(T \phi_{\xi} 
+\ds{1 \over 2} \psi_{\xi} \psi_{\xi} 
- \varepsilon_{\xi} \tp_{\xi} e 
- \ds{1 \over 2} e \tilde{\phi}_{\xi} e
- \ds{1 \over 2} \varepsilon_{\xi} R \varepsilon_{\xi})
\nonumber
\\
Z_1^3 (\xi) \!\!\! &=& \!\!\! -( \psi_{\xi} \phi_{\xi} 
- \varepsilon_{\xi} \tilde{\phi}_{\xi} e 
- \ds{1 \over 2} \varepsilon_{\xi} \tp_{\xi} \varepsilon_{\xi})
\nonumber
\\
Z_0^4 (\xi) \!\!\! &=& \!\!\! -\ds{1 \over 2} (\phi_{\xi}\phi_{\xi}
- \varepsilon_{\xi} \tilde{\phi}_{\xi} \varepsilon_{\xi} )
\, . 
\label{zny}
\end{eqnarray}

Last, we consider the Maxwell sector. It 
represents a topological 
Yang-Mills theory with Abelian gauge group 
which entails that the observables are generated 
by the $4$-form $F_{\ab} F_{\ab} $.

By combining all of these results, one concludes that the 
most general elements of the equivariant 
cohomology for topological gravity are obtained by considering 
appropriate products of the expressions
 constructed in the gravitational and 
Maxwell sectors.

\paragraph{Case $k=1$:}
This case can be treated along the same lines while starting from 
the closed $2$-form 
$W_2^0 \equiv \varepsilon^{ab} R_{ab}$: the generalized $2$-form 
\begin{equation}
\label{obs2d}
\hat W \equiv \varepsilon^{ab} \hat R _{ab} =
W_2^0 + W_1^1 (\xi) + W_0^2 (\xi)
\end{equation}
then satisfies $\hat d \hat W =0$
(i.e. $\cas \hat W = -d \hat W$) and involves the polynomials 
\begin{eqnarray}
\label{obs2db}
W_1^1 (\xi) \!\!\! & = &\!\!\! \varepsilon^{ab} (\tp _{ab} +i_{\xi} R_{ab})
\\
W_0^2 (\xi) \!\!\! & = &\!\!\! \varepsilon^{ab} (\tilde{\phi} _{ab} 
+i_{\xi} \tp _{ab} + \ds{1 \over 2} i_{\xi} i_{\xi} R_{ab})
\, .
\nonumber
\end{eqnarray}
From $\hat d (\hat W)^n =0$ for $n=1,2,\dots$
and from the expansion 
$(\hat W)^n = w_0^{2n}+ w_1^{2n-1} + 
w_2^{2n-2} $, 
one obtains 
more general representatives of the 
equivariant cohomology algebra
in two dimensions~\cite{verlinde, becchi}:
\begin{eqnarray}
w_0^{2n} \!\!\! & = &\!\!\! [W_0^2 (\xi ) ]^n 
\nonumber 
\\
w_1^{2n-1} \!\!\! & = &\!\!\! n [W_1^1  (\xi ) ][W_0^2 (\xi ) ]^{n-1}
\\
w_2^{2n-2} \!\!\! & = &\!\!\! n [W_2^0 (\xi ) ] [W_0^2 (\xi ) ]^{n-1} 
+ \ds{1 \over 2} \, n (n-1)  
[W_0^2 (\xi ) ] ^{n-2}  [W_1^1 (\xi ) ]^2 
\, .
\nonumber
\end{eqnarray}
There is no topological invariant related to the torsion
in two dimensions~\cite{zanelli}.
In the Maxwell sector, the basic invariant is given by 
the $2$-form $F_{\ab}$.

%%%%%%%%%%%%%%%%%%%%%%%%%%%%%%%%%%%%%%%%%%%%%%%%%%%%%%%%%%%%%%%%
\subsection{Second order formalism}
 
We only discuss the gravitational sector since the Maxwell 
sector is not modified.

If we require the torsion form to vanish, the connection 
becomes a function of the vielbein (and its inverse)
which is now the only independent
field in the gravitational sector:
\begin{equation}
 \label{omee}
 \omega_{abc} = \ds{1 \over 2} (P_{abc} +P_{bca}-P_{cab})
 \, , \quad {\rm with} \ \; 
 {P^a}_{bc} \equiv (\pa_{\mu} e_{\nu} ^a - \pa_{\nu} e_{\mu} ^a)
 e_b ^{\mu}  e_c ^{\nu}
 \, .
 \end{equation}
For consistency, one also has to require 
the $\cas$-variation of the torsion to vanish. 
By virtue of equation (\ref{sto}), this implies that $\tilde{\psi}$ is no
longer an independent ghost, rather it is a function of the ghost $\psi$ 
and the vielbein: 
\begin{equation}
   \label{ptps}
\tilde \psi ^a _{\ b} e^b  = D\psi ^a 
\, . 
 \end{equation}
Note that relation (\ref{omee}), as well as the solution
of  (\ref{ptps}) with respect to $\tilde \psi$,
again rely on the assumption that the vielbein is invertible. 

In summary, in the gravitational sector, 
the basic variable is the vielbein $e$ 
and we have ghosts $\xi,\, c, \, \psi$ as well as ghosts for ghosts 
$\varphi$ and $\tilde{\varphi}$ with the nilpotent $\cas$-variations
 \begin{equation}
\label{brs1b}
\begin{array}{lll}
\cas e =  \LL_\xi e  - ce + \p  \ ,
& \quad &
\cas \p =  \LL_\xi \p - c\p 
- \LL_\vf e + \tvf e  
\\
\cas c =  \LL_\xi c - c^2 +  \tvf \ , 
& \quad  &
\cas \tvf =  \LL_\xi \tvf - [c,\tvf] -\LL_\vf c  
\\
\cas \xi =  \xi^2 + \varphi 
\ , 
& \quad  &
\cas  \varphi = [\xi, \varphi]
\, .
\end{array}
\end{equation}
Henceforth, the only difference with the first order formalism consists
of the fact that 
$\omega,\, \tilde{\psi}$ and their $\cas$-variations
\begin{equation}
\label{some}
\cas \om =  \LL_\xi \om  -Dc  +  \tp 
\ , \quad 
\cas \tp =  \LL_\xi \tp - [c,\tp] 
- \LL_\vf \omega -D\tvf 
\end{equation}
are no longer independent expressions, but merely 
consequences of $T=0, \, \cas T =0$ and of the variations
(\ref{brs1b}).
A fortiori, the Lorentz part of 
the equivariant cohomology 
is not modified up to the fact that 
$\omega$ and $\tilde{\psi}$ have the explicit form 
(\ref{omee}) and (\ref{ptps})
in terms of $e$ and $\psi$.
Some of the results of the metric approach 
simply 
follow by combining 
the vielbein fields into a metric tensor
$g_{\mu \nu} \equiv \eta_{ab} e_{\mu}^a e_{\nu}^b$,
see appendix A.

%%%%%%%%%%%%%%%%%%%%%%%%%%%%%%%%%%%%%%%%%%  
%%%%%%%%%%%%%%%%%%%%%%%%%%%%%%%%%%%%%%%%%% 
%%%%%%%%%%%%%%%%%%%%%%%%%%%%%%%%%%%%%%%%%% 

\section{Superspace approach}\label{Superspace-approach}

%\section{Symmetries} 
%%%%%%%%%%%%%%%%%%%%%%%%%%%%%%%%%%%%% 

After introducing superspace and 
the geometric objects that it supports,
we derive the symmetry algebra of topological gravity
from a few simple transformation laws in superspace.
As before, we do not need to specify 
the space-time dimension for the discussion
of symmetries.

%%%%%%%%%%%%%%%%%%%%%%%%%%%%%%%%%%%%%%%%%%%%%%%%%%%%%%%
\subsection{ Supersymmetry and superspace}\label{superspace}  

{\em Rigid supersymmetry}
is defined by an odd generator  $Q$ satisfying 
the Abelian superalgebra $Q^2=0$. Field theoretic representations are
given by 
doublets and singlets, and they are 
readily obtained from a 
superspace construction: we  extend the 
$d$-dimensional 
space-time manifold by a single Grassmann  
variable $\theta$ so as to obtain a superspace parametrized 
by local coordinates $(x, \theta)$. 
Then,
a {\em superfield} is, by  definition,
a function on superspace,
\begin{equation}
\label{sf}
F(x,\theta) = f(x) + \theta f'_\th(x)
\, ,
\end{equation}
transforming under an infinitesimal supersymmetry transformation as
\eq
QF = \dth F\ ,
\eqn{def-sf}
which yields
the following action of $Q$ on the component fields:
\eq
Qf = {f'_\th}\ ,\quad Q{f'_\th}=0 \ .
\eqn{comp-susy-transf}
In expression \equ{sf}, the component field 
$f$
has the same Grassmann  grading as 
$F$
while its superpartner 
${f'_\th}$ 
has the opposite one. 
 We assign a ``supersymmetry ghost-number" 
({\em SUSY-number} or {\em SUSY-charge} for short) 
to all fields 
and variables: this charge is defined by assigning 
the value $-1$ to the variable $\theta$ and, quite generally,
an upper or lower $\th$-index on a field corresponds to a 
charge $-1$ or $+1$, respectively.
The generator $Q$ raises the SUSY-number by one unit. 

A {\em $p$-superform} admits the expansion  
\eq 
\hO_p (x,\theta) = \dsum{k=0}{p} \, \OM_{p-k}(x,\theta) \; (d\theta)^k
\, , 
\eqn{s-form} 
where $\OM_{p-k}$ has $k$ $\theta$-indices that we did not
spell out. 
The components $\OM_q (x,\theta)$ of the $p$-superform are $q$-forms
whose coefficients are superfields:
\[ 
\OM_q (x,\theta) = \frac{1}{q!} \, \OM_{\m_1...\m_q}(x,\theta)  
\; dx^{\m_1} \cdots dx^{\m_q}
= \om_q (x) + \theta \om'_{q\th} (x)
\, .
\]
In the previous expression and in the following, 
the wedge product symbol is always omitted.
Moreover, we will adhere to the notational conventions 
considered in the previous expressions:
functions or forms on ordinary space-time are 
denoted by small case letters, superfields (or space-time forms 
having superfields as coefficients) 
by upper case letters and super $p$-forms 
with $p \geq 1$ 
(i.e.
$p$-forms in  superspace with superfields as coefficients) 
by upper case letters with a ``hat''. 

A {\em supervector field} has the form 
$\hXI (x,\theta)= \Xi^\m (x,\theta) \pa_\m + \Xi^\theta (x,\theta)\dth
 \equiv \Xi ^M \pa_M$ 
where $M$ denotes a  supercoordinate index (i.e. $M=\m$ or
$M=\theta$). 
The  graded Lie bracket of two vector fields $\hXI_1$ and 
$\hXI_2$
is again a vector field whose components are given by 
\[
[ \hXI_1, \hXI_2 ] ^M = \Xi_1^N \pa_N \Xi_2^M \pm
\Xi_2^N \pa_N \Xi_1^M \ ,
\]
with a plus sign if both $\hXI_1$ and $\hXI_2$ 
have odd ghost-number, and a minus sign otherwise.

We now proceed to introduce the standard differential operators
in superspace. 
The exterior derivative is given by  
$\hd = d+d\theta \dth$ with $d = dx^\m \pa_\m$.
We have the relations  
$0= \hd^{\, 2} = d^{\, 2} = (d\theta \dth)^2 
= [d,d\theta \dth]$ where the bracket 
$[\cdot , \cdot]$ denotes the graded  commutator.
The Lie derivative  $\hLiexi$ with respect to 
the supervector field $\hXI$
acts on a superform according to 
 $\hLiexi = [i_\hXI , \hd \, ]$
(where $i_\hXI$ denotes the inner product operation) 
%and it acts on a supervector field $\hXI ^{\prime}$ according to 
%$\hLiexi  \hXI ^{\prime} = [\hXI, \hXI ^{\prime} \,]$.  
%In any case, 
and we have the graded 
commutation relation 
$ [\LL_{\hXI_1}, \LL_{\hXI_2}] = \LL_{ [\hXI_1 , \hXI_2 ]}$. 

A local, infinitesimal {\em supersymmetry transformation} 
 is given by a 
$x$-dependent 
translation of the $\theta$-variable, i.e.
$\theta \to \theta + \e ^{\theta} (x)$. Thus, it is a supercoordinate
transformation generated by the vector field 
$\e ^{\theta} (x) \pa_{\theta}$. The latter acts in superspace 
by virtue of the Lie derivative, e.g. on a superfield: 
\begin{equation} 
   \label{susyt}
\delta_{\e} F (x,\theta) 
= \e ^{\theta} (x) \pa_{\theta} F (x,\theta)
\, . 
\end{equation} 
The induced variations of the component fields read as
\begin{equation} 
\delta_{\e } f(x) = \e ^{\theta} (x)  {f'_\th}(x)
\quad ,\quad 
\delta_{\e }{f'_\th}(x)=0
\, . 
\label{def-Q} 
\end{equation} 
  Obviously, the rigid supersymmetry transformations 
$\delta F = \e \, QF$ with $\e$ constant 
%defined in \equ{def-sf} 
 represent a special case.

%%%%%%%%%%%%%%%%%%%%%%%%%%%%%%%%%%%%%%%%%%%% 
%\subsection{Topological gravity in superspace} 
%\subsubsection{Fields and symmetries within the BRST formalism}

\subsection{Fields and symmetries}\label{fields&symm}

The basic variables in the gravitational sector 
of the theory are the 
connection super $1$-forms $\hO^a{}_b(x,\th)$ 
associated to local Lorentz transformations and the vielbein 
super $1$-form $\hE^a(x,\th)$.
We do not introduce 
superforms with $\th$-indices, 
$\hO^{\th}_{\ \th} (x,\th)$
or $\hE^\th (x,\th)$.
In fact, $\hO^{\th}_{\ \th}=0$ since the action of the 
Lorentz algebra on scalars is trivial and 
$\hE^\th$ only transforms linearly and solely under supercoordinate
transformations,
henceforth there is no obstruction for its vanishing.
We shall however come back to this point in 
subsection~\ref{altern-appr}.
In the Maxwell sector, the basic variable is the 
connection super $1$-form $\hA$ associated 
to local $U(1)$ transformations.

Within the BRST formalism, the parameters of infinitesimal 
symmetry transformations are turned into ghost fields
(having a ghost-number $1$): 
thus, we have the Lorentz 
and Maxwell ghosts  $C^a{}_b(x,\th)$ 
and $U(x, \th )$ 
which are superfields
and the superdiffeomorphism ghost $\hXI$ which is a 
supervector field. 
The connection $\hO$ and ghost $C$ both take their values in the 
Lorentz algebra, i.e.
$\hO_{ab}=-\hO_{ba}$ and $C_{ab} = - C_{ba}$. 
For the ghost vector field  $\hXI$,  
it is convenient to introduce the notation
\[
\hXI^2 \equiv {1 \over 2} \, [ \hXI, \hXI ]
\; , \qquad {\rm i.e.} \ \; 
(\hXI^2)^M = \Xi^\m\pa_\m \Xi^M + \Xi^\th\dth \Xi^M
\, ,
\]
in terms of which we can write 
$ (\hLiexi)^2 = \LL _{\hXI^2}$.

%%%%%%%%%%%%%%%%%%%%%%%%%%%%%%%%%%%%%%%%%%%%%%%%%%%%%%%%
\subsection{BRST transformations in superspace}

%Within the BRST formalism, the parameters of local, infinitesimal 
%symmetry transformations are turned into ghost fields.
%The latter have ghost-number $g=1$ while the basic fields
%appearing in the invariant action (the connection) have
%a vanishing ghost-number.
 The Grassmann parity of an object is given by the 
parity of its {\it total degree} 
which is now 
defined as the sum $p+g+s$ of 
its form degree $p$,  ghost-number $g$ and SUSY-number $s$.
All commutators and  brackets are assumed to be graded 
according to this grading.  

We collect 
all symmetry transformations in the {\em superspace BRST 
transformations} 
which can be written in the following way, using 
 obvious matrix notation like  $\hE$ for $\hE^a$ and 
$\hO$ for $\hO^a{}_b$:   
\eq\ba{lll}
\cas \hE =     \hLiexi\hE -C \hE \ ,
\quad \quad &
\cas \hO  =  \hLiexi\hO  - \hat D C \ , 
\quad \quad & 
\cas \hA = \hLiexi \hA - \hat d U 
\\
\, \cas \hXI = \hXI^2 \ ,
\quad \quad &
\cas C =  \hLiexi C - C^2 \ ,  
\quad \quad &
\cas U =  \hLiexi U
\  .
\ea\eqn{super-BRS}
Here, $\hat D C \equiv \hd C + [ \hO ,C]$ and 
the given $\cas$-operator is nilpotent.

%\subsection{Horizontality conditions}
We note~\cite{bbg} that the transformations laws \equ{super-BRS} may be
deduced from horizontality conditions
involving the torsion and curvature superforms
\eq
\hat T \equiv \hd\hE + \hO \hE
\, ,\quad 
\hat R \equiv  \hd\hO + \hO^2
\, ,\quad 
\hat F \equiv  \hd \hA
\, .
 \eqn{super-curv}
Indeed, let us introduce the {\it extended superforms} 
$\EE\equiv \hE ,\, \OO \equiv \hO  +C,\, \AA \equiv  \hA + U$ 
and the {\it extended differential}
$\D \equiv  \hd + \cas - \LL_\hXI$.
The nilpotency requirement for $\D$ is equivalent 
to the transformation law $\cas \hXI = \hXI^2$. 
The extended torsion and curvature superforms
\eq
\TT \equiv  \D\EE+\OO \EE
\, ,\quad
\RR \equiv  \D\OO+\OO^2
\, ,\quad
\FF \equiv  \D\AA 
\eqn{ext-curv}
then satisfy the extended Bianchi identities
\eq
\D\TT+\OO \TT-\RR \EE =0
\, ,\quad 
\D\RR+[\OO,\RR]=0
\, ,\quad 
\D\FF =0
\, .
\eqn{ext-bianchi}
The BRST transformations \equ{super-BRS}
now result from the {\em horizontality conditions}
\[
\TT=\hat T 
\, , \quad  
\RR=\hat R 
\, , \quad  
\FF =\hat F
\]
and substitution of these conditions into 
the extended Bianchi identities (\ref{ext-bianchi}) directly 
yields the transformation laws of the 
torsion  and curvature superforms:
\eq
\cas \hat T = \LL_\hXI\hat T - C\hat T 
\, ,\quad
\cas \hat R = \LL_\hXI\hat R - [C,\hat R]
\, ,\quad
\cas \hat F = \LL_\hXI\hat F
\, .
\eqn{brs-curv}

%%%%%%%%%%%%%%%%%%%%%%%%%%%%%%%%%%%%%%%%%%%%%%%
\subsection{Projection to component fields}\label{Proj-to}

\subsubsection{General gauge}
In order to obtain the space-time BRST transformations, 
we introduce the superfield components of superforms,  
\begin{eqnarray}
\hE^a \!\!\!&=&\!\!\!   E^a(x,\th) + d\th \, E^a_\th(x,\th)
\qquad 
\ \ \, {\rm with} \ \; E^a = dx^{\mu} \, E^a_{\mu}
\nonumber
\\
\label{e1}
\hO ^a{}_b \!\!\!&=&\!\!\!   \OM^a{}_b(x,\th) + d\th \, 
\OM_\th^a{}_b(x,\th)
\quad \; \; \ \ {\rm with} \ \;  \OM^a{}_b  = dx^{\mu} \, \OM^a_{\mu b}
\\
\hA  \!\!\!&=&\!\!\!   A (x,\th) + d\th \, A_\th(x,\th)
\qquad 
\, \ \quad {\rm with} \ \; A = dx^{\mu} \, A_{\mu}
\, ,
\nonumber
\end{eqnarray}
as well as the space-time components of the latter:
\begin{equation}
\label{stc}
\ba{ll}      
E^a(x,\th)  =   e^a(x)+\th 
\psi_{\theta} ^a(x)
\, ,  \quad 
&E^a_\th(x,\th) = \chi_\th^a(x)+\th\f_{\th\th}^a(x) 
\es
\OM^a{}_b(x,\th)  = \om^a{}_b(x)+\th\tp_\th^a{}_b(x) 
 \, ,  \quad 
&\OM_\th^a{}_b(x,\th) =  \tchi_\th^a{}_b(x)
+\th\tf_{\th\th}^{\ a}{}_b(x) 
\es
A (x,\th) = \ab (x)+\th \eta_{\th} (x)
\, ,  \quad 
&A_\th(x,\th) =  
\sigma_{\th}  (x) +\th {t}_{\th \th} (x) 
\, .
\ea
\end{equation}
Similarly, we define the component fields of the ghost superfields: 
\begin{equation}
\ba{ll}  
\Xi^\m(x,\th)  =  \xi^\m(x) +
 \th{\xi}_\th^{\prime \m}(x)  
\, , \quad &
C^a{}_b(x,\th) = 
c^a{}_b(x) +  \th {c}_\th^{\prime \, a}{}_{ b}(x)
\\
\Xi^\th (x,\th) =   \ep^\th (x) + \th 
{\epp} (x)
\, , \quad & 
U (x,\th) = u(x) +  \th 
u_{\th} ^{\prime} (x)
\, . 
\ea
\end{equation}

In the sequel, 
  we will omit the indices labeling space-time fields
  in order to simplify the notation and 
we will use the short-hand notation $\xi (x) \equiv \xi ^{\mu} (x)
\pa_{\mu}$ and $\xip (x) \equiv \xi ^{\prime \mu} (x) \pa_{\mu}$.

From equations (\ref{super-BRS}) it follows that 
the {\em BRST transformations of space-time fields}
take the following form 
(where $Dc \equiv dc +[\omega , c]$
denotes the Lorentz covariant derivative):
\begin{eqnarray}
\cas e  \!\!\!&=&\!\!\! \LL_\xi e  - ce + \ep \p - d\ep \, \chi
\nonumber \\
\cas \p  \!\!\!&=&\!\!\! \LL_\xi \p - \LL_{\xip} e   - c\p + c'e  
- \epp \p -d\epp \, \chi   - d\ep \, \f 
\nonumber \\
\cas \chi  \!\!\!&=&\!\!\! \LL_\xi \chi - i_{\xip}{}e  - c\chi 
 + \ep \f - \epp \chi 
\label{BRST-e} 
\\
\cas \f  \!\!\!&=&\!\!\! \LL_\xi \f - \LL_{\xip} \chi 
- i_{\xip}{}\p   
- c\f + c'\chi  - 2 \epp \f 
\nonumber 
\end{eqnarray}
\begin{eqnarray}
\cas \om \!\!\!&=&\!\!\! \LL_\xi \om 
-Dc
+ \ep \tp - d\ep \, \tchi 
\nonumber \\
\cas \tp \!\!\!&=&\!\!\! \LL_\xi \tp - \LL_{\xip}\om   - [c,\tp]  
-Dc' 
- \epp \tp - d\epp \tchi -d\ep \, \tf
\nonumber \\
\cas \tchi \!\!\!&=&\!\!\! \LL_\xi \tchi - i_{\xip}{}\om  - [c,\tchi] -
c'  + \ep \tf - \epp \tchi 
 \label{BRST-omega} 
 \\
\cas \tf \!\!\!&=&\!\!\! \LL_\xi \tf - \LL_{\xip} \tchi - i_{\xip}{}\tp 
 - [c,\tf]  + [c',\tchi]  - 2 \epp \tf 
\nonumber 
\end{eqnarray}
%%%%%%%%
\begin{eqnarray}
\cas c \!\!\!&=&\!\!\!  \LL_\xi c - c^2 +\ep c'
\nonumber \\
\cas c' \!\!\!&=&\!\!\! \LL_\xi c' - \LL_{\xip} c - [c,c'] - \epp c'
\label{BRST-ghosts} 
\\
\cas \xi \!\!\!&=&\!\!\! \xi^2 + \ep \xip 
\, , \qquad \quad  
\cas \xip = [\xi,\xip ] - \epp \xip 
\nonumber \\
\cas \ep \!\!\!&=&\!\!\! \LL_\xi \ep + \ep \epp
\, , \qquad  
\ \cas \epp  = \LL_\xi \epp - \LL_{\xip} \ep 
\, . 
\nonumber 
\end{eqnarray}
%%%%%%%%%%%%%
\begin{eqnarray}
\cas \ab  \!\!\!&=&\!\!\! \LL_\xi \ab  -du 
+ \ep \eta - d\ep \, 
{\sigma}
\nonumber \\
\cas \eta  \!\!\!&=&\!\!\! \LL_\xi \eta - \LL_{\xip} \ab  
-du ^{\prime}
- \epp \eta -d\epp \, 
{\sigma}
- d\ep \, t 
\nonumber \\
\cas {\sigma}  \!\!\!&=&\!\!\! \LL_\xi {\sigma} - 
i_{\xip}{} \ab - u ^{\prime}
 + \ep {t} - \epp {\sigma} 
\nonumber \\
\cas {t}  \!\!\!&=&\!\!\! \LL_\xi {t} - 
\LL_{\xip} {\sigma}
 - i_{\xip}{}\eta   - 2 \epp {t}
\label{BRST-a} 
\\
\cas u \!\!\!&=&\!\!\!  \LL_\xi u +\ep u'
\, , \qquad  
\cas u' =  \LL_\xi u' - \LL_{\xip} u  - \epp u'
\, .
\nonumber 
\end{eqnarray}
 We note that  these $\cas$-variations 
describe eight local symmetries, 
par\-am\-etriz\-ed by the ghosts 
$\xi ,\, \e, \, c, \, u$ and $\xi' ,\, \e', \, c', \, u'$.
The first four ones are the diffeomorphism, local supersymmetry, 
local Lorentz and local Maxwell 
transformations
whereas the last four ones 
may be called vector supersymmetry,
$R$- (or Fayet) transformations and supergauge  transformations.
As we shall see in the next subsection, one may, if one wishes, 
gauge fix the three local invariances 
parametrized by $\xi' ,\, c', \, u'$
in an algebraic way. 
In addition, the positive SUSY-numbers can be traded 
for positive 
ghost-numbers 
by rescaling fields with 
appropriate powers of $\e$, 
the consequence being that  
the parameters $\e$ and $\e'$ disappear
from the BRST transformations.

%%%%%%%%%%%%%%%%%%%%%%%%%%%%%%%%%%%%%%%%%%% 
\subsubsection{Wess-Zumino gauge}\label{wzsg}

%\subsubsection{ Topological gravity with the graviphoton} 

Besides the physically relevant fields and symmetries, 
the superfield formalism generally introduces 
some additional fields 
and symmetries which can be eliminated 
in an algebraic way 
by imposing supergauge conditions of   Wess-Zumino (WZ) type.
%which can be solved algebraically.  
In the present case,
the %{\em WZ supergauge} 
{\em WZ gauge} is defined by the choices 
\eq 
\chi = 0
\quad , \quad 
\tchi=0 
\quad , \quad 
{\sigma}=0 \, .
\eqn{wz-condition} 
 and it corresponds to the gauge-fixing of the local invariances 
parametrized by 
 the ghosts  $\xip$,  $c'$ and $u'$. 
 In fact, by virtue of equations \equ{BRST-e}, \equ{BRST-omega}
and \equ{BRST-a}, 
 the $\cas$-invariance of the choices \equ{wz-condition} 
 requires the conditions  
\eq 
\ep \f - i_{\xip} e =0
\quad , \quad 
\ep \tf - i_{\xip}\om - c'=0
\quad , \quad  
\ep {t} - i_{\xip} \ab - u' =0 
\, . 
\eqn{wz-stability} 
The latter allow us to eliminate the ghosts  $\xip$,  $c'$ and 
$u'$,  
\eq\ba{c} 
{\xi}^{\prime \m} = \ep \vf^\m
\quad ,\quad 
c'=\ep \tvf 
\quad ,\quad 
u'=\e \tau 
\, ,  
\ea\eqn{elimin} 
where $\vf^\m$, $\tvf$ and $\tau$  are defined by 
\eq 
\vf^\m = \f^a e_a^\m
\quad ,\qquad 
\tvf =  \tf-i_\vf\om
\quad ,\qquad 
\tau =  {t} - i_\vf \ab 
\, . 
\eqn{vec-fi} 
Here, $( e_a^\m )$ denotes the inverse vielbein,
i.e. the vielbein $(e^a_\m)$ is assumed to be invertible
at this point.  
If we consider $\xi^{\prime a} \equiv  \xi^{\prime \mu} e^a_\m$,
the first equations in (\ref{elimin}) and (\ref{vec-fi}) 
can be rewritten as 
$\xi^{\prime a} = \ep \f^a$ and $0= \f^a - i_\vf e ^a$,
respectively. Thus, each of the three expressions
appearing in equations (\ref{wz-condition}),(\ref{elimin}) 
and (\ref{vec-fi}) have the same form. 
 
By substituting the WZ gauge choices \equ{wz-condition} and their 
stability conditions
\equ{elimin} into the $\cas$-variations   
(\ref{BRST-e})-(\ref{BRST-a}), we find 
the BRST transformations in the WZ gauge.
Since the WZ gauge choices do not affect 
diffeomorphisms, Lorentz and Maxwell transformations,
we will only display the other contributions to the 
BRST transformations, i.e.
the parts parametrized by $\e$ and $\e '$.
For any space-time field $f$ with SUSY-number $\alpha_f$,
the $\e '$-variation reads as 
\begin{equation}
\label{fayet}
\delta_{\e '} f = \alpha_f\, \e ' f
\, , 
\end{equation}
very much like Fayet's $R$-transformation 
in ordinary (i.e. Poincar\'e) supersymmetric field theory.
The local supersymmetry transformations read as 
\begin{equation}
  \label{delvar} 
  \begin{array}{lllll}
\delta_{\e}  e = \e \psi 
\, , &\quad &
\delta_{\e}   \p  = -\e ( \LL_\vf e - \tvf e )
-2(d\e) (i_{\varphi} e) 
\, , &\quad &
\delta_{\e}  \xi = \ep ^2 \vf
\\
\delta_{\e}   \om  =  \e \tp 
\, , &\quad &
\delta_{\e}  \tp   = -\e  ( \LL_\vf \omega + D\tvf )
-2(d\e) (\tvf +  i_{\varphi} \omega)
\, , &\quad &
\delta_{\e}   c  = \e^2 \tvf
\\
\delta_{\e}  \ab =  \e \eta   
\, , &\quad &
\delta_{\e}  \eta = - \e ( \LL_{\vf} \ab + d  \tau ) 
-2(d\e) (\tau +  i_{\varphi} \ab) 
\, , &\quad &
\delta_{\e}  u =  \ep ^2 \tau 
\end{array}
\end{equation}
and $\delta_{\e} \e' = -\e \LL_\vf \e$. 
These results coincide in parts with the 
on-shell expressions obtained in references~\cite{spence,anselmi}
by twisting the on-shell version of $N=2$ 
euclidean supergravity\footnote{A comparison 
of both the field content and symmetries before and after
the twisting of an extended (rigid or local) supersymmetric
field theory shows that the operations of twisting and 
of reduction to the mass-shell (i.e. elimination 
of auxiliary fields) commute with each other.
We wish to thank B.~Spence for kindly 
illustrating this point to us.}.
    
In order to obtain the {\em shift symmetries} of 
topological gravity, one has to absorb 
the parameter $\e (x)$ of local supersymmetry
into the fields, just as one 
does in topological Yang-Mills theory for 
the constant parameter $\e$ of rigid supersymmetry,
e.g. see reference~\cite{bcglp}. 
More precisely, we absorb all $\th$-indices of the fields 
(which have been explicitly 
displayed in expressions (\ref{stc})) 
by rescaling these fields with appropriate powers of $\e \equiv \e ^{\th}$.
Since $\e$ has SUSY-number $-1$ and ghost-number $1$,
this rescaling amounts to assigning positive ghost-numbers
to these fields rather than positive SUSY-numbers.
Thus, let us redefine the variables according to 
\begin{equation}
\begin{array}{lllll}
\p _0 =\e\p \ ,
& \quad & 
\tp _0 =\e\tp \ , 
& \quad & 
\eta _0 =\e\eta
\\
\vf _0 =\e^2\vf \ ,
& \quad & 
\tvf _0 =\e^2\tvf \ , 
& \quad & 
\tau _0 =\e^2\tau
\ ,
\end{array}
\label{field-baulieu}
\end{equation}
without modifying the basic fields $e, \omega, \ab$
and the ghosts $\xi, c, u$. Then, the 
{\em BRST transformations in the WZ  gauge} take 
the form
\begin{equation}
  \label{wz-BRST-e'} 
  \begin{array}{lll}
\cas e  = \LL_\xi e  - ce +  \p_0 \ ,
&\quad &
\cas \xi = \xi^2 + \vf_0 
\\
\cas \p_0  = \LL_\xi \p_0  - c\p_0   -  \LL_{\vf_0 } e + \tvf_0  e  \ ,
&\quad &
\cas \vf_0   = [\xi, \vf_0 ] 
\\
&\quad &
\\
\cas \om  = \LL_\xi \om  -Dc + \tp_0  \ ,
&\quad &
\cas c  = \LL_\xi c - c^2 +  \tvf_0 
\\
\cas \tp_0   = \LL_\xi \tp_0  - [c,\tp_0 ] - \LL_{\vf_0 } \omega - D\tvf_0 \ ,
&\quad &
\cas \tvf_0   = \LL_\xi \tvf_0  - [c,\tvf_0 ] - \LL_{\vf_0 } c 
  \end{array}
\end{equation}
and 
\begin{equation}
\label{wz-BRST-a'} 
\begin{array}{lll}
\cas \ab  = \LL_\xi \ab - du 
+ \eta_0 \ ,
&\quad &
\cas u   = \LL_\xi u + \tau_0  
\\
\cas \eta_0   = \LL_\xi \eta_0  - \LL_{\vf_0 } \ab  - d\tau_0 \ ,
&\quad &
\cas \tau_0   = \LL_\xi\tau_0  - 
\LL_{\vf_0 } u
\, .
\end{array}
\end{equation}
%\label{wz-BRST-a'} 
%\label{wz-BRST-omega'}
%\eqn{wz-BRST-ghosts'} 
Supersymmetry is now realized as a rigid symmetry, 
as usual for the shift supersymmetry of topological 
field theories.
Remarkably enough, the transformation laws 
(\ref{wz-BRST-e'})(\ref{wz-BRST-a'}) coincide with 
those obtained in equations (\ref{brs1}),
i.e. those of reference~\cite{bautan1}.
The parameter $\ep$ has disappeared from 
these $\cas$-variations,
because it has been absorbed into the fields
so as to define new fields with vanishing SUSY-number.
Consequently, the parameter $\e '$ parametrizing the 
SUSY-number symmetry according to eq.(\ref{fayet})
does not occur either in these transformation laws.
To be more precise, $\e$ and $\e '$ only appear in 
$\cas \e$ and $\cas \e '$ which variations can simply
be omitted since they decouple from the others.

%%%%%%%%%%%%%%%%%%%%%%%%%%%%%%%%%%%%%%%%%%%%%
\subsection{Alternative approach}\label{altern-appr}

Interestingly enough, the BRST algebra
(\ref{wz-BRST-e'})(\ref{wz-BRST-a'})
can also be obtained by starting from different fields
and symmetries in superspace.
More precisely, let us discard the 
$U(1)$ superconnection $\hat A$ as well as the associated 
ghost $U$, and let us supplement the super $1$-form 
$\hat E ^a$ with a $\th$-component $\hat E ^{\th}$
transforming as 
$\cas \hE^\th = \hLiexi\hE^\th$. 
In other words, we are now
considering the complete superspace vielbein matrix:
\eq
\lp\ba{ll}
E^a_\m
\quad & E^a_\theta\es
E^\th_\m\quad & E^\theta_\theta
\ea\rp \, .
\eqn{vielbein}
The expansion of $\hat E ^{\th}$ reads as 
\[
 \hE^\th =   E^\th(x,\th) + d\th \, E^\th_\th(x,\th)
\qquad 
\ {\rm with} \ \; E^\th = dx^{\mu} \, E^\th_{\mu}
\]
and 
\[
E^\th(x,\th) = \ab ^\th(x)+\th \eta (x)
\, ,  \quad 
E^\th_\th(x,\th) =  \sigma_{\th}^{\th} (x) +\th {t}_{\th} (x) 
\, .
\]
Here, the space-time components of $E ^{\th}$
and $E^\th_\th$ have been denoted by the same letters as 
the components of $A$ and $A_{\th}$ 
in expressions (\ref{stc}), except for the fact that these components 
now carry an extra upper index $\th$.
By projecting the superspace BRST transformations
to space-time components, one gets the 
$\cas$-variations \equ{BRST-e}-(\ref{BRST-a}) with $u=0=u'$ 
in the last set of equations.

The WZ gauge choices are again given by
$\chi =0 = \tilde{\chi}$ (see eqs.(\ref{wz-condition})), 
but the condition ${\sigma}\equiv {\sigma}_{\th} =0$
is now replaced by $\sigma_{\th}^{\th} =1$.
As before, the stability of the gauge choices 
$\chi =0 = \tilde{\chi}$
under $\cas$-variations implies 
${\xi}^{\prime \m} = \ep \vf^\m$ and $c'=\ep \tvf$. 
The stability of the condition $\sigma_{\th}^{\th} =1$
now leads to 
\eq
\ep t_\th - i_{\xip} \ab^\th - \e'=0\ ,
\eqn{stab-sigma}
and thus allows us to eliminate the ghost $\e'$ 
by virtue of the relation
$\e'=\e \tau $ with 
$\tau \equiv  {t}_\th - i_\vf \ab^\th $.
If we  substitute all of these expressions 
into the transformation rules \equ{BRST-e}-(\ref{BRST-a}) 
and subsequently perform the field redefinitions
(\ref{field-baulieu}), we again 
obtain  the BRST transformations \equ{wz-BRST-e'}, as well 
as 
\equ{wz-BRST-a'} with $u$  replaced by $\e$.
Henceforth, the $\e$-transformations (which only concern
the field $\ab$ 
and its partners $\eta_0 $, $\tau_0 $) 
should no longer be interpreted as local supersymmetry
transformations, but rather as $U(1)$ gauge transformations.
Accordingly, the space-time 
field $\ab$ is to be viewed as Maxwell 
potential, i.e. as graviphoton field.
Of course, this reinterpretation of the variables
$\e,\, \ab, \, \eta_0 , \, \tau_0$ 
requires a change of statistics for each of them. 
Since all of these fields carry an (upper) index $\th$,
our reinterpretation is tantamount to dropping this index,
i.e. shifting their SUSY-number from one to zero.
There is no obstruction to this shift, because the ghost $\e'$
parametrizing the SUSY-number symmetry has been eliminated
by virtue of the WZ gauge choices. 

\paragraph{Remark:}     
The stability of a 
non-vanishing value for $\sigma_{\th}^{\th}$
is ensured by condition \equ{stab-sigma}
which determines the ghost $\e'$ in terms of the ghost
$\e$. 
Since the variable $\sigma_{\th}^{\th}$ represents the 
lowest component of the superfield
$E_{\th} ^{\th}$, this condition
(together with the invertibility of 
the vielbein matrix $(e^a_{\mu})$
which is related 
to the stability of the gauge choice $\chi=0$)
ensures that the  supervielbein matrix \equ{vielbein}
is invertible.
Let us stress 
that this invertibility 
has only to be imposed 
in the Wess-Zumino gauge. 
Degenerate supervielbeins 
%with singularities 
may well appear in a general gauge. 
This situation is somewhat reminiscent of the fact that 
the invertibility problem does not manifest itself 
in $3$-dimensional  quantum gravity if the latter is
expressed  as a topological theory of Chern-Simons or
of $BF$ type~\cite{witten-3d-grav, bf}.
%This situation is very similar with quantum gravity
%in three dimensions expressed as a topological theory of the
%Chern-Simons or of the $BF$ type~\cite{witten-3d-grav}.

It is puzzling that two approaches involving 
different fields, symmetries and gauge choices lead to 
space-time results of the same form.
To elucidate this point, we consider the $\cas$-variations
(\ref{BRST-a}) which have been obtained 
 in a general gauge, in the case where 
Maxwell transformations were included at the superspace 
level:
\begin{eqnarray*}
\cas \ab  \!\!\! &=& \!\!\!
- du - (d\e) \sigma +\dots \\
\cas \eta  \!\!\! &=& \!\!\! - du' - (d\e ') \sigma  +\dots \\
(\cas - \LL_{\xi} )\sigma   \!\!\! &=& \!\!\! - u' - \e ' \sigma + \e t - 
i_{\xi '} \ab
\, .
\end{eqnarray*}
Thus, the parts of the $\cas$-variations in superspace
that are parametrized by 
$U \equiv u + \th u '$ and 
$\Xi ^{\th} \equiv \e + \th \e '$ 
 yield the same space-time results
for $\cas \ab, \, \cas \eta$ and $\cas \sigma$ up to a 
factor $\sigma$: for every term in $u$ or $u'$, respectively, there 
is analogous term in $\e$ or $\e'$, multiplied 
by $\sigma$.  
In the superspace approach based on $\hat A$ and $U$, the gauge 
choice $\sigma =0$ implies
\[
\begin{array}{l}
\cas \ab  = - du  +\dots \\
\cas \eta  = - du'  +\dots
\, , \qquad {\rm with} \ \, 
u' =\e (t -  i_{\vf} \ab)
\, .
\end{array}
\]
By contrast, in the approach based on $\hat E ^{\th}$ 
and $U=0$ (i.e. $u=0=u '$), the gauge 
choice $\sigma =1$ implies
\[
\begin{array}{l}
\cas \ab  = - d\e   +\dots \\
\cas \eta  = - d\e'  +\dots
\, , \qquad {\rm with} \ \, 
\e' = \e (t -  i_{\vf} \ab)
\, .
\end{array}
\]

%%%%%%%%%%%%%%%%%%%%%%%%%%%%%%%%%%%%%%%%%%%%%%%%%%%%%%%%%%%%%%%%
\subsection{Observables}

Superspace expressions for the observables 
related to the curvature 
may be obtained by viewing the theory as a topological 
gauge theory associated to the Lorentz group and to a $U(1)$ group:
the methods developed for topological Yang-Mills theories
in reference~\cite{bcglp} can then be applied.
They also allow
us to obtain space-time expressions for the observables
in a general gauge (and not just in the WZ gauge).

%%%%%%%%%%%%%%%%%%%%%%%%%%%%%%%%%%%%%%%%%%%%%%%%%%%%%%%%%%%%%%%%
%%%%%%%%%%%%%%%%%%%%%%%%%%%%%%%%%%%%%%%%%%%%%%%%%%%%%%%%%%%%%%%% 
%%%%%%%%%%%%%%%%%%%%%%%%%%%%%%%%%%%%%%%%%%%%%%%%%%%%%%%%%%%%%%%%

\section{Remarks on the gauge fixing}

The complete Lagrangian for topological gravity can be 
constructed 
by gauge fixing the shift symmetry (characterizing a 
topological invariant) by virtue of a condition which
localizes the path integral so as to describe a moduli
space of interest,  e.g. see ref.~\cite{bautan1}.
Examples of such gauge choices which can be imposed onto the Lorentz
connection are the {\em flatness} condition $R_{\mu \nu} =0$ (which is
admissible in any space-time dimension), the {\em half-flatness} or 
{\em self-duality} condition $R_{\mu \nu}^- =0$ (in four dimensions)
or the condition of {\em constant scalar curvature} 
(in two dimensions)~\cite{birmrak,oda,stora}.

On a four-dimensional Riemannian manifold with $SU(2)$ holonomy,
one has the following remarkable result concerning 
self-duality~\cite{eguhan, bautan1}. The curvature 
two-form $R^a_{\ b}$ 
of a torsionless connection $\omega^a_{\ b}(e)$ satisfies 
the self-duality condition $R_{ab}^- =0$
(where $X^-_{ab} \equiv {1 \over 2} (X_{ab} - {1 \over 2}
\varepsilon_{abcd} X_{cd})$)
if and only if $\omega^a_{\ b}(e)$ is self-dual, i.e. 
$\omega_{ab}^-(e)=0$.
In this respect, we note that 
the gauge group $SO(4)$ is locally given by
$SU(2)_+ \otimes  SU(2)_-$ and  that the condition $R_{ab}^- =0$
is $SO(4)$-invariant, whereas the condition $\omega_{ab}^-(e)=0$
is only $SU(2)_+$-invariant.
The self-dual part $\omega_{ab}^+(e)$ transforms like a connection
and the $SU(2)$ holonomy 
corresponds to a reduction of the $SO(4)$-frame bundle 
to a  $SU(2)$-bundle. Accordingly, one 
expects a restricted action of local orthonormal 
transformations on fields like the one 
encountered in reference~\cite{spence}. 

Alternatively, the complete Lagrangian may be 
obtained by twisting 
$N=2$ euclidean supergravity~\cite{anselmi, spence} on a 
Riemannian four-manifold with $SU(2)$ holonomy.
Indeed such a manifold
admits two covariantly constant chiral spinors
that can be used to perform the twist of the gravitinos and of the 
parameters of supergravity transformations
so as to give rise to shift transformations parametrized by
a variable $\varepsilon (x)$.
(This was recently done in detail using an on-shell
formulation~\cite{spence}). 
Our discussion in subsection \ref{wzsg} shows that the variable
$\varepsilon (x)$ has to be absorbed in an appropriate way into the
fields if one wants to cast the shift transformations 
into a standard form and to compare with the models constructed
by gauge fixing a topological invariant.

The twist of supergravity transformations not only gives rise
to local supersymmetry transformations
parametrized by the scalar $\varepsilon (x)$, but  
also to a vector and a tensor supersymmetry
for which on-shell expressions have been given in reference
\cite{spence}. By contrast to the global vector supersymmetry
transformations encountered in topological Yang-Mills
theory~\cite{brand},  the local vector supersymmetry of 
topological gravity  does not leave invariant the 
fundamental fields
(i.e. the vielbein and graviphoton) 
and therefore appears to act
non-trivially on the topological invariant from which the complete
Lagrangian originates by gauge fixing. Thus, this symmetry may be
more restrictive  for the perturbative renormalization of 
topological gravity than it is for topological Yang-Mills theory.

\section{Concluding comments}

We have shown that the graviphoton field $\ab_\m$ can be implemented in the
superspace approach in two different ways, namely 
using an independent Abelian 
superconnection as in subsection \ref{Proj-to} or, 
maybe more geometrically, 
using a complete superspace vielbein as in
subsection \ref{altern-appr}. 
Although both implementations involve different 
local symmetries, 
they turn out to be equivalent in the sense that they lead to the 
same space-time 
BRST algebra, once the supergauge is fixed according to suitable
Wess-Zumino type conditions.

It is worth mentioning once more that the vielbein matrix 
does not need to be invertible in superspace, 
although the formulation in the
Wess-Zumino gauge (that corresponds to the various formulations 
considered 
in the literature), necessitates an invertible vielbein, 
i.e. 
a metric which is nonsingular at every space-time point. Thus,
the theory written in superspace in a general supergauge might 
%be more general than the one defined in the Wess-Zumino gauge.
have further significance than the   
 one defined in the Wess-Zumino gauge.

%%%%%%%%%%%%%%%%%%%%%%%%%%%%%%%%%%%%%%%%%%%%%%%%%%%%%%%%%%%%%%%% 

\vskip 1.2truecm
 
{\bf \Large Acknowledgments}
 
\vspace{3mm}

%\nopagebreak 

It is a great pleasure to thank 
M. Blau, F. Delduc, B. Spence, R. Stora 
and G. Thompson for valuable discussions. Olivier Piguet 
thanks the members of the 
Institut de Physique Nucl\'eaire of the University of Lyon 
for a very kind invitation as Visiting Professor for two periods during
which a large part of this work has been done.

%%%%%%%%%%%%%%%%%%%%%%%%%%%%%%%%%%%%%%%%%%%%%%%%%%%%%%%%%%%%%%%%

\appendix 
\section{Appendix:   Metric approach}

We will only discuss the gravitational sector since the 
Maxwell sector can be treated as in the vielbein formalism. 

\noindent
{\bf Notation:}
The metric formulation involves tensor fields, 
e.g. the metric tensor field 
$g = g_{\mu \nu} dx^{\mu} \otimes  dx^{\nu} $.
Their variation under an infinitesimal change of coordinates
generated by the vector field $w = w^{\mu} \partial_{\mu}$
is given by the Lie derivative ${\cal L}_{w}$
as acting on generic tensor fields, e.g.
\begin{equation}
{\cal L}_{w} g_{\mu \nu} \equiv ({\cal L}_{w} g)_{\mu \nu} =
w^{\rho} \partial_{\rho} g_{\mu \nu} 
+ (\partial_{\mu} w^{\rho}) g_{\rho \nu} 
+ (\partial_{\nu} w^{\rho}) g_{\rho \mu} 
\, .
\end{equation}
In particular, the action of ${\cal L}_{w}$ on a 
vector field $\varphi = \varphi ^{\mu} \partial_{\mu}$ 
is the Lie bracket (\ref{grlbr}):
${\cal L}_{w} \varphi^{\mu} = [w, \varphi ]^{\mu}$.

The metric tensor field 
can also be viewed as a $0$-form with values in the 
covariant rank-two tensors. 
Along the same vein, 
the collection $(\G_{\m\s}^\n)$  of Christoffel symbols
may be regarded as a matrix-valued $1$-form,
$\G^\n{}_{\s}$ $=$ $\G_{\m\s}^\n dx^\m$. 
Thus, these geometric quantities can be acted upon 
by the linear operator 
$l_w \equiv [i_w , d ] = i_w d + (-1)^{[w]} di_w$,
e.g. 
\begin{eqnarray}
l_{w} g_{\mu \nu}  
\!\!\! & = & \!\!\!
w^{\rho} \partial_{\rho} g_{\mu \nu}
\nonumber 
\\
(l_w \Gamma) _{{\bf \mu} \, \cdot} ^{\, \cdot} 
\!\!\! & = & \!\!\!
w^{\rho} \partial_{\rho} \Gamma_{{\bf \mu} \, \cdot} ^{\, \cdot} 
+ ( \pa_{{\bf \mu}} w^{\lambda}) \Gamma_{\lambda \, \cdot} 
^{\, \cdot}  
\ .
\label{def-l}
\end{eqnarray}
Note that the operator $l_w$ and the Lie derivative 
${\cal L}_{w}$ act in the same way on forms which do not carry
extra curved space indices $\mu , \nu, \dots$, like the forms appearing
in the vielbein formalism (which carry extra tangent space
indices $a,b, \dots$).
This should be kept in mind when comparing the results below 
with those presented in the main body of the text. 

\noindent
{\bf Fields:}
As in the second order formalism, we
eliminate the Lorentz connection $\omega$ 
 in terms of the vielbein $e$ 
 by requiring the torsion to vanish. 
The metric tensor  
 given by $g_{\mu \nu} = \eta_{ab} e^a_{\mu} e^b_{\nu}$
 can then be considered as the only independent variable. 

\noindent
{\bf Symmetries:}
The BRST algebra reduces to
the following well known form~\cite{mp}: 
\begin{equation}
\label{brsg}
\begin{array}{lll}
\cas g =
\LL _\xi g  
+ \Psi
\, , & \quad & 
\cas \Psi =  
\LL _\xi \Psi - \LL _\vf g
\\
\cas \xi  =  \xi^2 + \varphi 
\, , & \quad & 
\cas \varphi = [\xi, \varphi]
\, .
\end{array}
\end{equation}
 Here, the symmetric tensor field $\Psi$ with components 
\begin{equation}
\label{compsi}
\Psi_{\mu \nu}  \equiv \eta_{ab}
( e^a_{\mu} \psi^b_{\nu} + e^a_{\nu} \psi^b_{\mu} ) 
= \psi _{\mu \nu} + \psi _{\nu \mu}  
\end{equation}
is to be viewed as a new variable that is usually 
referred to as {\em gravitino} field.

We note that the diffeomorphisms can be completely 
decoupled by introducing 
the operator 
$\check{\cas} \equiv \cas - \LL _{\xi}$
which satisfies $\check{\cas}^2 =  - 
\LL _\vf
$.
The variations of $g, \, \Psi$ and $\vf$,  
as given by (\ref{brsg}), then read as 
\begin{equation} 
 \check{\cas} g =   \Psi\quad , \quad 
 \check{\cas} \Psi =   - 
\LL _\vf
g \quad , \quad 
 \check{\cas}  \vf = 0  \, .
   \end{equation}
%and $\check{\cas} \xi =  - \xi^2 + \varphi$.
Thus, one has a close analogy with topological Yang-Mills theory
where $\check{\cas}$ corresponds to the SUSY-generator $\tilde Q$  
that is nilpotent up to an infinitesimal gauge transformation.

From equations (\ref{brsg}), it follows that the Christoffel symbols
describing  the Levi-Civita connection transform as 
\begin{equation}
\cas \Gamma_{{\bf \mu} \sigma} ^{\rho} =
(l _{\xi} \Gamma) _{{\bf \mu} \sigma} ^{\rho} 
+ \nabla_{\mu}  ( \pa_{\sigma} \xi^{\rho})
+  \tilde{\Psi} ^{\rho}_{{\bf \mu} \sigma}
\ .
\label{brsg1}
\end{equation}
Here,  $l_{\xi} \Gamma $  is given by (\ref{def-l})
and $\nabla_{\mu}$  denotes 
the covariant derivative with respect to the 
Levi-Civita connection, i.e. 
$\nabla_{{\bf \mu}} v^{\rho}_{\ \sigma} \equiv 
\pa_{{\bf \mu}}  v^{\rho}_{\ \sigma}
+ \Gamma_{{\bf \mu} \lambda} ^{\rho} v^{\lambda}_{\ \sigma}
- \Gamma_{{\bf \mu} \sigma} ^{\lambda} v_{\ \lambda}^{\rho}$, 
while the components of the 
third rank tensor $\tilde \Psi $ are defined by 
\begin{equation}
\label{tilp}
\tilde{ \Psi} ^{\rho}_{{\bf \mu}\sigma} \equiv 
 {1 \over 2} \, 
(  \nabla_{{\bf \mu}} \Psi^{\rho}_{\ \sigma}  
 + \nabla_{\sigma} \Psi^{\ \rho}_{{\bf \mu}}
 -  \nabla^{\rho} \Psi_{{\bf \mu} \sigma} ) 
\ . 
\end{equation}
Since the variable $\pa_{\sigma} \xi^{\rho}$ appearing 
in equation (\ref{brsg1}) does not define a tensor field,
the expression  $\nabla_{\mu}(\pa_{\sigma} \xi^{\rho})$
only represents a convenient notation.
As a matter of fact, 
the right-hand-side of equation (\ref{brsg1}) can also 
be written in terms of the Lie derivative acting 
on tensor fields~\cite{bertlmann}:
\begin{equation}
\cas \Gamma_{{\bf \mu} \sigma} ^{\rho} =
\LL _{\xi} \Gamma _{{\bf \mu} \sigma} ^{\rho} 
+ \partial_{\mu}  ( \pa_{\sigma} \xi^{\rho})
+  \tilde{\Psi} ^{\rho}_{{\bf \mu} \sigma}
\ .
\label{brsg7}
\end{equation}
However, just as for expression (\ref{brsg1}), this only represents 
a convenient notation since the Christoffel symbols do not 
define a tensor field.

%%%%%%%%%%%%%%%%%%%%%%%%%%%%%%%%%%%%%%%%%%%
\subsection{Horizontality conditions}

By using matrix notation, we can derive the 
BRST algebra (\ref{brsg})(\ref{brsg1}) from an  
horizontality condition.
To do so, let us introduce the matrix-valued forms 
\begin{eqnarray}
\underline{\Gamma}  \!\!\!&=&\!\!\! 
\underline{\Gamma} _{\mu} dx^{\mu} 
\, , \qquad \qquad \, {\rm with} \ \; 
\underline{\Gamma}_{\mu} = ( \Gamma_{\sigma \mu} ^{\rho} )
\label{mvf}
\\
\underline{R}  \!\!\!&=&\!\!\! {1 \over 2} \, 
\underline{R}_{\mu \nu} dx^{\mu} dx^{\nu} 
\, , \quad \ {\rm with} \ \ \, 
 \underline{R}_{\mu \nu} = ( R^{\rho} _{\ \sigma\mu \nu} )
\ ,
\nonumber 
\end{eqnarray}
where $\underline{R}$  denotes the
curvature $2$-form associated to the Levi-Civita connection.
The connection forms 
$\omega$ and $\underline{\Gamma}$ are related by a formal
gauge transformation involving the matrix of 
vielbein fields  $E \equiv (e_{\mu} ^a)$:
\begin{equation}
\omega_{\mu} = E \underline{\Gamma}_{\mu} E ^{-1} + E \partial_{\mu} E 
^{-1}
\, .
\label{fgt}
\end{equation}
Accordingly, the curvature $2$-forms 
$R$ and $\underline{R}$ associated to $\omega$ and 
$\underline{\Gamma}$,
respectively, are related by a similarity transformation: 
$R = E \underline{R} E ^{-1}$.

Let us now consider the {\em generalized fields} 
\cite{ader} 
\begin{equation}
\begin{array}{lll}
   \hat{\underline{\Gamma} }  \equiv
{\rm e}^{i_{\xi}} \, ( \underline{\Gamma} + \underline{v} ) 
=  \underline{\Gamma} + \underline{v} + i_{\xi} \underline{\Gamma} \ ,
& \quad \ {\rm with} &  
\ \underline{v} = ( v_{\ \sigma} ^{\rho} ) \equiv 
(\pa_{\sigma} \xi ^{\rho} )
\\
\hat{\underline{R} }   \equiv 
\hat d \, \hat{\underline{\Gamma} } +
\hat{\underline{\Gamma} }^2 \ ,  & \quad \ {\rm with} & \
\hat d = d+\cas  \, . \nonumber 
\end{array}
\end{equation}
%\underline{v} = ( v_{\ \sigma} ^{\rho} ) 
%\, ,  \ \ \qquad \quad \quad {\rm with} \ \; 
%v_{\ \sigma} ^{\rho} = \pa_{\sigma} \xi ^{\rho}
By construction, $\hat{\underline{R} } $
satisfies the generalized Bianchi identity
$0= \hat{\nabla} \hat{\underline{R} } 
\equiv \hat d \, \hat{\underline{R} }
+ [ \hat{\underline{\Gamma} } ,  \hat{\underline{R} } ]$.
Since $(\underline{v} + i_{\xi} \underline{\Gamma})^{\rho}
 _{\
\sigma} = \nabla_{\sigma} \xi ^{\rho}$, 
we also consider the covariant derivative 
$\nabla_{\sigma} \vf^{\rho} $ as well as 
the combination of covariant derivatives (\ref{tilp})
which describes the shift of $\underline{\Gamma}$:
\begin{eqnarray}
\underline{\tilde{\Phi}}  \!\!\!&=&\!\!\!
( \tilde{\Phi} _{\ \sigma} ^{\rho} )
\, , \ \qquad  \, \ \ \ {\rm with} \ \; 
\tilde{\Phi} _{\ \sigma} ^{\rho} = 
\nabla_{\sigma} \vf^{\rho} 
\nonumber 
\\
\label{pre}
\underline{\tilde{\Psi}}  \!\!\!&=&\!\!\!
\underline{\tilde{\Psi}}_{\mu} dx^{\mu} 
\, , \qquad \quad {\rm with} \ \; 
\underline{\tilde{\Psi}}_{\mu} 
= ( \tilde{\Psi} ^{\rho} _{\sigma \mu} ) 
\, .
\end{eqnarray}

The {\em horizontality condition} then reads as 
\begin{equation}
\label{mhor}
\hat{\underline{R} } =  {\rm e}^{i_{\xi}} \, (  \underline{R} 
+ \underline{\tilde{\Psi}} + \underline{\tilde{\Phi}} )
\end{equation}
and we can proceed as in subsection \ref{hhh}
to derive the BRST transformations. 
(Instead of assuming the fields $\underline{\tilde{\Psi}}_1^1$
and $\underline{\tilde{\Phi}}_0^2$ appearing in 
(\ref{mhor}) to be explicitly given by (\ref{pre}), we could also 
assume them to be undetermined. 
The consistency 
of the resulting BRST transformations
with the known $\cas$-variations of $\xi$ and
$g$ and with the expression for 
the Christoffel symbols in terms of the metric, then 
{\em implies} the relations (\ref{pre}).) 

 Thus, we use the operatorial relation (\ref{oprel}), 
the definitions (\ref{brs2})
which are part of the basic algebra (\ref{brsg}),
as well as a change of variables that is analogous to (\ref{cvar}):  
\[
\underline{\tilde{\Phi}} \to \underline{\tilde{\varphi}}
\equiv  \underline{\tilde{\Phi}} - i_{\vf}
\underline{\Gamma}  \ , \qquad 
{\rm i.e.} \ \; \tilde{\varphi} ^{\rho} _{\ \sigma} =
\pa_{\sigma} \vf^{\rho}
\, . 
\]
Thereby, the $\cas$-variations following from the 
expansion of (\ref{mhor})
and of the Bianchi identity 
$\hat{\nabla} \hat{\underline{R} } =0$ 
take the form 
\begin{equation}
\begin{array}{lll}
\label{brs6}
\cas \underline{\Gamma} = l_\xi \underline{\Gamma} 
- \nabla \underline{v} +  \underline{\tilde{\Psi}} \nonumber 
\ , &\quad & 
\cas \underline{\tilde{\Psi}} =
l_\xi  \underline{\tilde{\Psi}}   
- [ \underline{v} ,  \underline{\tilde{\Psi}} ] 
- l_{\varphi}  \underline{\Gamma}
 -\nabla  \underline{\tilde{\varphi}}
 \\
\cas \underline{v} = l_\xi  \underline{v} -  
\underline{v}^2 + \underline{\tilde{\varphi}} 
\ , &\quad & 
\cas \underline{\tilde{\varphi}} 
=  l_\xi  \underline{\tilde{\varphi}} 
- [ \underline{v}, \underline{\tilde{\varphi}} ]
- l_{\varphi}  \underline{v}
\\
\cas \xi =  \xi^2 + \varphi 
\ , &\quad & 
\cas  \varphi = [\xi, \varphi]
\end{array}
\end{equation}
and 
\begin{equation}
\label{brs7}
\cas \underline{R} =  l_\xi \underline{R} - 
[\underline{v},\underline{R}]  
-\nabla \underline{\tilde{\Psi}}
\ ,
\end{equation}
where $\nabla  \underline{v} \equiv d\underline{v} 
 + [\underline{\Gamma} ,\underline{v}]$.
Due to relation (\ref{fgt}), 
the BRST algebra (\ref{brs6}) of the metric
approach has exactly the same form as 
the BRST algebra (\ref{some})(\ref{brs1b}) of the vielbein approach
(which entails that the BRST variations (\ref{brs6}) are nilpotent). 
In fact, when passing from the vielbein formalism
to the metric approach, a tangent space index that is acted upon
by the Lorentz parameter $c^{ab}$ 
(with $\cas c^{ab} = \tilde{\varphi}^{ab} + \dots$) becomes a 
curved space index that is acted upon by diffeomorphisms
in the disguise of the parameter 
$v^{\mu}_{\ \nu} = \partial_{\nu} \xi^{\mu}$ 
(with 
$\cas v^{\mu}_{\ \nu} =  \tilde{\varphi}^{\mu} _{\ \nu} +\dots
= \partial_{\nu} \varphi ^{\mu} +\dots$). 

We note that the symmetry algebras of the prepotential $g$, 
as given by equations (\ref{brsg}), and of the 
 potential $\underline{\Gamma}$, as given by (\ref{brs6}),
are consistent with each other and that these 
symmetry algebras have the same structure:
\begin{equation}
\begin{array}{lcl}
\cas g = \delta_{\xi} g + \Psi \, , & \quad & 
\cas \Psi = \delta_{\xi} \Psi - \delta_{\varphi} g \\
\cas \underline{\Gamma} = \delta_{\xi} \underline{\Gamma} 
+ \underline{\tilde{\Psi}} \, , & \quad & 
\cas  \underline{\tilde{\Psi}} = \delta_{\xi} \underline{\tilde{\Psi}}
 - \delta_{\varphi}  \underline{\Gamma}
\, .
\end{array}
\end{equation}

%The potential $\underline{\Gamma}$ and its field strength 
%$\underline{R}$ are functions of the prepotential $g_{\mu \nu}$ 
%and their  transformation laws, as given by (\ref{brs6})(\ref{brs7}),
%are consistent with those of $g_{\mu \nu}$.
%Similarly, the ``field strengths'' $\underline{\tilde{\Psi}}, \, 
%\underline{v}$ and  $\underline{\tilde{\varphi}}$ of,
%respectively,     $\Psi , \, \xi$ and $\vf $
%are consistent with the transformation laws 
%of the latter, as given by equations (\ref{brsg}). 

%The ghost $\underline{v}$ describing infinitesimal ``Lorentz 
%transformations'' and the field 
%$\underline{\tilde{\phi}}$ parametrizing the shift of 
%$\underline{v}$ are, respectively, functions of the 
%diffeomorphism ghost vector field 
%$\xi^{\mu} \pa_{\mu}$ and of its shift $\varphi^{\mu}\pa_{\mu}$.
%Thus, their $\cas$-variations  represent consequences of
%the ones of $\xi$ and $\varphi$.

%%%%%%%%%%%%%%%%%%%%%%%%%%%%%%%%%%%%%%%%%%%%%%%%%%%%%%%%%%

\subsection{Comparison with the vielbein approach} 
 
Let us compare the variables appearing, respectively,  
in the metric approach 
and in the second order formalism.
 The shift transformations of fields 
 are symbolized by a vertical arrow:
\[
\qquad \qquad \qquad \qquad \qquad
\mbox{Metric approach}
\qquad \quad 
\mbox{Second order formalism}
\]
\[
\begin{array}{lcccc}
\mbox{Basic fields:}  &\quad &  g_{\mu \nu}  & &            
   \\
   &&\downarrow   & &             
\\   
\mbox{Ghosts:}&&\psi_{\mu \nu}  &, &  \xi^{\mu}  
   \\
  &&              &  &     \downarrow  
\\               
\mbox{Ghosts for ghosts:}    & &            &  & \vf^{\mu}  
\end{array}
\qquad \quad 
\begin{array}{ccccc}
   e_{\mu}^a  & &   & &          
   \\
   \downarrow   & &    & &               
\\   
\psi_{\mu}^a  &, &  \xi^{\mu}    &, & c^{ab} 
\\
                &  &     \downarrow  &  &     \downarrow   
\\  
                &  & \vf^{\mu} & , & \tvf^{ab}  \ . 
\end{array}
\]
The basic field $e_{\mu}^a$ of the second order formalism
involves a Lorentz index, which implies that a Lorentz ghost 
and the corresponding ghost for ghost appear 
as independent variables,
in addition to those that are present in the metric approach. 
In particular, the ghost for ghost 
$\tilde{\varphi}^{ab}$ appears 
in the variation $\cas  \psi_{\mu}^a$ and thereby 
in the observables of the vielbein formalism, 
 though it can only 
appear in those of the metric approach
under the disguise of the dependent variable 
$\tilde{\varphi}^{\rho}_{\ \sigma} = 
\partial_{\sigma} \varphi^{\rho}$, 
see next subsection. 
One expects that the observables 
in these different approaches are cohomologically equivalent,
i.e. that they only differ by $\cas$- and $d$-exact terms,
just as the gravitational anomaly in Einstein gravity 
can manifest itself under different disguises
(Lorentz anomaly or diffeomorphism anomaly, as well as 
Weyl or chirally split anomaly in two 
dimensions)~\cite{bertlmann, klt}.

%%%%%%%%%%%%%%%%%%%%%%%%%%%%%%%%%%%%%%%%%%%%%%%%%%%%%%%%%%

\subsection{Observables}

In view of the formal gauge transformation (\ref{fgt}), 
one would expect that the expressions 
for the observables in the second order formalism have exactly 
the same {\em form} as those in the metric approach.
Yet, this is not quite true as we will see in the following.

Let us denote the observables in the metric approach by 
$\cm _d^0$, $\cm _{d-1}^1, \dots, \cm _0^d$
so as to distinguish them from those of the vielbein 
formalism denoted by 
$W _d^0$, $W _{d-1}^1, \dots, W _0^d$.
The polynomials 
$\cm _{d-k}^k$
satisfy descent equations 
that are analogous to equations (\ref{des4d}) which correspond 
to the special case $d=4$.
Of course, the topological invariant 
$\cm _d^0 (g_{\mu \nu})$ coincides with the topological invariant 
$W_d^0 (e_{\mu}^a)$
since the metric $g_{\mu \nu}$ can 
be expressed in terms of the vielbein fields $e_{\mu}^a$.
Furthermore, the polynomial 
$\cm _{d-1}^1 (g_{\mu \nu}, \Psi_{\mu \nu}, \xi^{\mu} )$ 
coincides with 
$W _{d-1}^1 ( e_{\mu}^a , \psi_{\mu}^a, \xi^{\mu} )$ 
by virtue of relations (\ref{compsi}) and (\ref{ptps}).
However, the polynomials of ghost-number $k\geq 2$, i.e. 
$\cm _{d-k}^k  (g_{\mu \nu}, \Psi_{\mu \nu}, \xi^{\mu} , \vf^{\mu})$,
do not depend on the same set of independent variables
as $W_{d-k}^k ( e_{\mu}^a , \psi_{\mu}^a, \xi^{\mu} ,  
\vf^{\mu}, \tilde{\varphi}^{ab})$.
And even if $\tilde{\varphi}^{ab}$ (or $\tilde{\phi}^{ab}$)
is viewed as the Lorentz analogue of 
$\tilde{\varphi}^{\mu}_{\ \nu} \equiv \partial_{\nu} 
\vf^{\mu}$ (or $\tilde{\Phi}^{\mu}_{\ \nu} \equiv \nabla_{\nu} 
\vf^{\mu}$), the polynomials of ghost-number
$k\geq 2$
do not quite have the same form: the expressions 
$\cm _{d-k}^k , \dots$, 
involve extra contributions which are not present in 
$W_{d-k}^k , \dots$.
We will see that the appearance of these terms can be drawn back 
to the 
shift transformations affecting the metric tensor which raises 
or lowers covariant indices.

\paragraph{Two-dimensional case:}

One starts from the $2$-form 
$\cm  ^0 _2 =  {\rm As}\, \underline{R} $
where $\underline{R}$ is the matrix-valued $2$-form 
defined in equations \equ{mvf}
and `${\rm As}$' the antisymmetric part of this matrix: 
\[
\cm  ^0 _2 = 
{\rm As}\,  \underline{R} \equiv  
 \sqrt{g} \, \varepsilon_{\rho \sigma}  
  \underline{R}^{\rho \sigma}
  =    {1 \over 2} \,  
 \sqrt{g} \, \varepsilon_{\rho \sigma}    
 R^{\rho \sigma} _{\ \ \mu \nu} \, dx^{\mu}  dx^{\nu} 
= {1 \over 2} \, \sqrt{g} \, \varepsilon_{\mu \nu} 
{\cal R} \, dx^{\mu}  dx^{\nu} 
 \, .
  \]
 Here, 
%$\eta_{\rho \sigma}  \equiv  \sqrt{g} \, \varepsilon_{\rho \sigma}$
  $g$ denotes the determinant of the metric tensor,
$\varepsilon_{\rho \sigma}$ the antisymmetric 
 tensor in flat space (which is just a numerical tensor
normalized by $\varepsilon_{12}=1$)
and ${\cal R}$ the curvature scalar. 

By laboriously solving the descent equations 
\[
\cas \cm ^0 _{2}  = - d \cm ^1 _{1} (\xi) \, , \quad 
\cas \cm ^1 _{1}  (\xi) = - d \cm ^2 _{0} (\xi) \, , \quad 
\cas \cm ^2 _{0} (\xi)= 0\, , 
\]
one finds the following
expressions~\cite{becchi},  which correspond to the 
Mum\-ford clas\-ses~\cite{bau-sin}:
\begin{eqnarray}
\cm ^1 _{1} (\xi) \!\!\!&=&\!\!\! 
 \sqrt{g} \, [ \, \varepsilon_{\mu \nu} 
 \,  \nabla^{\nu} \Psi_{\ \rho} ^{\mu} 
 + \varepsilon_{\mu \rho}  \, 
\xi ^{\mu} {\cal R} \, ]  \, dx^{\rho} 
 \label{d2e}
 \\
 \cm ^2 _{0} (\xi) \!\!\!&=&\!\!\! 
 \sqrt{g} \, \varepsilon_{\mu \nu}
 \, [ \,  
\nabla^{\mu} \varphi ^{\nu} 
- {1 \over 4} \, \Psi^{\mu \rho} \Psi _{\rho}^{\ \nu} 
 - \xi ^{\rho} \nabla^{\nu} \Psi_{\ \rho} ^{\mu} 
 + {1 \over 2} \, \xi ^{\mu}  \xi ^{\nu}  {\cal R} \, ] 
   \, .
\nonumber
\end{eqnarray}
Using the two-dimensional identity 
``$0= \varepsilon_{\mu \nu} V_{\rho}\, +$  cyclic permutations 
of indices'', the term involving a 
derivative of $\Psi^{\mu}_{\ \rho}$ may also be 
expressed in terms of the traceless part of 
the symmetric tensor $\Psi$:
\[
 \varepsilon_{\mu \nu} 
\, \nabla^{\nu} \Psi_{\ \rho} ^{\mu} =
-  \varepsilon_{\mu \nu}  \nabla_{\rho} 
(\Psi^{\nu \rho} - g^{\nu \rho} \Psi^{\sigma}_{\ \sigma} )
\, .
\]
In reference \cite{stora}, 
the results (\ref{d2e}) have been obtained 
by applying the 
mathematical techniques of equivariant cohomology,
thereby justifying earlier calculations
and discussions along these lines~\cite{myers2, wu}.

Alternatively, one could try to proceed as 
in subsection~\ref{observables} 
(see equations (\ref{obs2d}) and (\ref{obs2db})),
i.e. consider the expansion
\begin{eqnarray}
\hat{\cm} 
 \!\!\!&\equiv &\!\!\!  
{\rm As}\,  \hat{\underline{R}}  
= {\rm e}^{i_{\xi}} \,  {\rm As}\, (  \underline{R}
+  \underline{\tilde{\Psi}} + \underline{\tilde{\Phi}} )
\label{essai} 
\\
 \!\!\!&=&\!\!\! 
{\rm As}\,  \underline{R} \, + \, 
{\rm As}\, (\underline{\tilde{\Psi}} 
+ i_{\xi} \underline{R} ) 
\, +  \, {\rm As}\,  
( \underline{\tilde{\Phi}} 
+  i_{\xi} \underline{\tilde{\Psi}} 
+ {1 \over 2}  i_{\xi} i_{\xi} \underline{R} )
 \, .
\nonumber 
\end{eqnarray}
The latter expressions have exactly the same form  
as the polynomials $W_2^0$, $W_1^1 (\xi)$ and $W_0^2 (\xi)$
appearing in 
eqs.(\ref{obs2d})(\ref{obs2db})
of the vielbein approach.
By spelling them out, one immediately finds the results
(\ref{d2e}) up to the quadratic term 
$ \Psi^{\mu \rho} \Psi _{\rho}^{\ \nu}$ 
that is present  in $\cm _0 ^2 (\xi) $. 
Such a term is generated from the variation
\[
\cas \Psi^{\mu}_{\ \nu} = - \Psi^{\mu \rho} \Psi _{\rho \nu}
+ \LL _{\xi} \Psi^{\mu}_{\ \nu} - g^{\mu \rho} \, 
(\LL _{\varphi} g_{\rho \nu})
\, ,
\]
i.e., it is due to the fact that the metric tensor, which 
raises or lowers indices, is subject to shift
transformations. 
This shows that the purely algebraic 
passage from ordinary to generalized fields and from 
the  ordinary differential $d$ to the  
generalized differential $\hat d = d + \cas$ is a subtle business
for topological models 
in the metric approach. A proper geometric 
treatment requires to 
extend the action of symmetries from the space-time manifold
to the infinite-dimensional space of all metrics, 
whence the use of global differential geometric machinery, 
see ref.~\cite{stora}.

In conclusion, we note that the two-dimensional metric tensor 
(and thus the two-dimensional  observables)
can equally well be parametrized
in terms of Beltrami differentials, see references~\cite{bau-sin, stora}.

%%%%%%%%%%%%%%%%%%%%%%%%%%%%%%%%%%
\paragraph{Four-dimensional case:}

One starts from the $4$-form 
\begin{equation}
\cm_4^0 = E^{\mu \rho} _{\lambda \chi} \  
\underline{R}^{\lambda}_{\ \mu} \, 
\underline{R}^{\chi}_{\ \rho} 
\, ,
\end{equation}
where 
\[
E^{\mu \rho} _{\lambda \chi} =
\left\{
\begin{array}{lcl}
\delta_{\lambda}^{\mu} \, \delta_{\chi}^{\rho} -
\delta_{\chi}^{\mu} \, \delta_{\lambda}^{\rho}
& \quad & \mbox{for the Pontryagin density} 
\\
&&
\\
{1 \over \sqrt{g} } \, 
\varepsilon^{\mu \nu \rho \sigma} \, g_{\nu \lambda}
\, g_{\sigma \chi} 
& \quad & \mbox{for the Euler density} 
\, ,
\end{array}
\right.
\]
i.e. 
\[
\cm_4^0 =
\underline{R}^{\mu \nu} \, 
\underline{R}_{\mu \nu} 
\qquad {\rm or} \qquad 
\cm_4^0 = 
{1 \over \sqrt{g} } \,
\varepsilon^{\mu \nu \rho \sigma} \,
\underline{R}_{\mu \nu} \, 
\underline{R}_{\rho \sigma} 
\, .
\]
The first descendant can  readily be 
obtained by expanding the generalized 
$4$-form $\hat {\cm} \equiv E^{\mu \rho} _{\lambda \chi} \  
\underline{\hat R}^{\lambda}_{\ \mu} \, 
\underline{\hat R}^{\chi}_{\ \rho} $ with respect to the ghost-number: 
\[
\cm_3^1(\xi) = 2 
E^{\nu \sigma} _{\mu \rho} \, 
\underline{R}^{\mu}_{\ \nu} \, 
[ \, {1 \over 2} \, (\nabla_{\sigma} \Psi^{\rho} _{\ \beta}
- \nabla^{\rho} \Psi^{\sigma \beta} ) + \xi^{\alpha}
R^{\rho} _{\ \sigma \alpha \beta} \, ] \, dx^{\beta}
\, . 
\]
For a determination and explicit expression of the other polynomials, 
we refer to the work~\cite{thuillier}.

Finally, we note that the Nieh-Yan $4$-form (\ref{nieh}), which  
yields the observables related to torsion, takes the form~\cite{ny}
\[
Z_4^0 = \ds{1 \over 4} \, \sqrt{g} \, \varepsilon^{\mu \nu \rho \sigma} 
\, (R_{\mu \nu \rho \sigma} -    \ds{1 \over 2} {T^{\lambda}}_{\mu \nu}
T_{\lambda \rho \sigma} ) \, dx^1 \dots dx^4        
\, .
\]

%%%%%%%%%%%%%%%%%%%%%%%%%%%%%%%%%%%%%%%%%%%%%%%%%%%%%%%%%%%%%%%
%%%%%%%%%%%%%%%%%%%%%%%%%%%%%%%%%%%%%%%%%%%%%%%%%
%\newpage 

%%%%%%%%%%%%%%%%%%%%%%%%%%%%%%%%%%%%%%%%%%%%%%%%%%

\begin{thebibliography}{99}



\bibitem{ewitten}
E.~Witten, ``Topological quantum field theory",
{\em Commun.Math.Phys.} {\bf 117} (1988) 353;

E.~Witten, ``Introduction to cohomological theories'',
{\em Int.J.Mod.Phys.} {\bf A6} (1991) 2775.

\bibitem{smolin}
L.~Smolin and A.~Starodubtsev,
``General relativity with a topological phase: 
an action principle'', 
{\tt hep-th/0311163};

A. Perez,  ``Spin foam models for quantum gravity'',
{\em Class.Quant.Grav.} {\bf 20} (2003) R43,
{\tt gr-qc/0301113};

T. Thiemann, ``Lectures on loop quantum gravity'', in 
{\it Quantum Gravity: From Theory to Experimental Search,}
D.J.W. Giulini, C. Kiefer and C. L\"ammerzahl, eds.,
Lecture Notes in Physics Vol.631 (Springer Verlag, 2003), 
{\tt gr-qc/0210094}; 

I.~Oda,
``A relation between topological quantum field theory 
and the Kodama state'', 
{\tt hep-th/0311149};

G.~Bonelli and A.M.~Boyarski, 
``Six dimensional topological gravity and 
the cosmological constant problem'', 
{\em Phys.Lett.} {\bf B490} (2000) 147, 
{\tt hep-th/0004058}.

\bibitem{qft}
E.~Witten, in ``Quantum Fields and Strings: A Course 
for Mathematicians, Vol.~2", P.~Deligne et al. (eds.),
(American Mathematical Society, 1999).



\bibitem{witten}
E.~Witten,
``Topological gravity,''
{\em Phys.Lett.}  {\bf B206} (1988) 601;

J.~M.~F.~Labastida and M.~Pernici,
``A Lagrangian for topological gravity and its BRST quantization,''
{\em Phys.Lett.} {\bf B213} (1988) 319;


R.~Brooks, D.~Montano and J.~Sonnenschein,
``Gauge fixing and renormalization in topological quantum field 
theory,''
{\em Phys.Lett.} {\bf B214} (1988) 91;


D.~Montano and J.~Sonnenschein,
``Topological strings,''
{\em Nucl.Phys.}  {\bf B313} (1989) 258.


\bibitem{montano}
D.~Montano and J.~Sonnenschein,
``The topology of moduli space and quantum field theory,''
{\em Nucl.Phys.}  {\bf B324} (1989) 348;


J.~M.~F.~Labastida, M.~Pernici and E.~Witten,
``Topological gravity in two dimensions,''
{\em Nucl.Phys.} {\bf B310} (1988) 611.


\bibitem{mp}
R.~Myers and V.~Periwal,
``Topological gravity and moduli space,''
{\em Nucl.Phys.} {\bf B333} (1990) 536.

\bibitem{myers2}
R.~Myers,
``New observables for topological gravity,''
{\em Nucl.Phys.} {\bf B343} (1990) 705.

\bibitem{myers1}
R.~Myers,
``On alternate formulations of topological gravity,''
{\em Phys.Lett.} {\bf B252} (1990) 365.

\bibitem{12a} 
R.~Myers and V.~Periwal,
``Invariants of smooth 4-manifolds from topological gravity,''
{\em Nucl.Phys.} {\bf B361} (1991) 290.


\bibitem{verlinde}
E.~Verlinde and H.~Verlinde,
``A Solution Of Two-Dimensional Topological Quantum Gravity'',
{\em Nucl.Phys.} {\bf B348} (1991) 457.


\bibitem{bau-sin}
L.~Baulieu and I.M.~Singer, 
``Conformally invariant gauge fixed actions for 2-d topological 
gravity'', 
  {\em Commun.Math.Phys.} {\bf 135} (1991) 253.

\bibitem{becchi}
C.~M.~Becchi, R.~Collina and C.~Imbimbo,
``On the semi-relative condition for closed (topological) strings'',
{\em Phys.Lett.}  {\bf B322} (1994) 79, 
{\tt hep-th/9311097};

C.~M.~Becchi, R.~Collina and C.~Imbimbo,
``A functional and Lagrangian formulation of two dimensional 
topological  gravity'',
in {\it Symmetry and Simplicity in Physics: 
a symposium on the occasion of Sergio Fubini's 65th Birthday,} 
W.M. Alberico and S. Sciuto, eds.  
(World Scientific, 1994), 
{\tt hep-th/9406096};

C.~M.~Becchi and C.~Imbimbo,
``Gribov horizon, contact terms and \v{C}ech- De Rham cohomology 
in 2D topological gravity'',
{\em Nucl.Phys.} {\bf B462} (1996) 571, 
{\tt hep-th/9510003};

C.~M.~Becchi and C.~Imbimbo,
``A Lagrangian formulation of 2-dimensional topological gravity and
\v{C}ech-De Rham cohomology,'' {\tt hep-th/9511156}.


\bibitem{nso}
A.~Nakamichi, A.~Sugamoto and I.~Oda,
``Topological four-dimensional self-dual gravity,''
{\em Phys.Rev.} {\bf D44} (1991) 3835.


\bibitem{oda}
I.~Oda,
``Topological four-dimensional gravity,''
{\em Progr.Theor.Phys.Suppl.} {\bf 110} (1992) 41.

\bibitem{wu}
S.~Wu,
``Appearance of universal bundle structure in 
four-dimensional topological
gravity'', {\em J.Geom.Phys.}  {\bf 12} (1993) 205.


\bibitem{stora}
R.~Stora, F.~Thuillier and J.~C.~Wallet,
``Algebraic structure of cohomological field theory models 
and equivariant cohomology'', in 
{\it Infinite dimensional geometry, non commutative geometry, 
operator algebras and fundamental interactions,}
Proceedings of the Caribbean Spring School of Mathematics 
and Theoretical Physics 
(Guadeloupe 1993), 
R. Coquereaux, M. Dubois-Violette, P. Flad, eds.  
(World Scientific, 1995).

\bibitem{thuillier} 
F. Thuillier, 
``Some remarks on topological 4d-gravity'', 
{\em J.Geom.Phys.} {\bf 27} (1998) 221,  
{\tt hep-th/9707084}.

\bibitem{mt}
M.~Menaa and M.~Tahiri,
``BRST-anti-BRST symmetry and observables for topological gravity,''
{\em Phys.Rev.}  {\bf D57} (1998) 7312.


\bibitem{perry}
M.~J.~Perry and E.~Teo,
``Topological conformal gravity in four dimensions,''
{\em Nucl.Phys.}  {\bf B401} (1993) 206, 
{\tt hep-th/9211063}.

\bibitem{anselmi}  
D. Anselmi and P. Fr\'e,
``Twisted N=2 supergravity as topological gravity 
in four dimensions'',
{\em Nucl.Phys.} {\bf B392} (1993) 401,  
{\tt hep-th/9208029}.

\bibitem{bautan1}
L.~Baulieu and A.~Tanzini,
``Topological gravity versus supergravity on manifolds 
with special  holonomy'',
{\em JHEP} {\bf 0203} (2002) 015,  
{\tt hep-th/0201109}.

\bibitem{spence}
P.~de Medeiros and B.~Spence,
``Four-dimensional topological Einstein-Maxwell gravity'', 
{\em Class.Quant.Grav.}  {\bf 20} (2003) 2075,  
{\tt hep-th/0209115}.


\bibitem{bfgrav}
H.Y.~Lee, A.~Nakamichi and T.~Ueno,
``Topological two form gravity in four dimensions'', 
{\em Phys.Rev.} {\bf D47} (1993) 1563,
{\tt hep-th/9205066};

M.~Abe, A.~Nakamichi and T.~Ueno,
``Gravitational instantons and moduli spaces 
in topological two form gravity'',
{\em Phys.Rev.} {\bf D50} (1994) 7323.

\bibitem{bautan2}
L.~Baulieu, M.~Bellon and A.~Tanzini,
``Eight-dimensional topological gravity and its correspondence 
with  supergravity'', 
{\em Phys.Lett.}  {\bf B543} (2002) 291,  
{\tt hep-th/0207020}; 

%\bibitem{bautan3}
L.~Baulieu, M.~Bellon and A.~Tanzini,
``Supergravity and the knitting of the Kalb-Ramond two-form 
in  eight-dimensional topological gravity'', 
{\em Phys.Lett.} {\bf B565} (2003) 211, 
{\tt hep-th/0303165};

%\bibitem{bautan4}
L.~Baulieu,
``Gravitational topological quantum field theory 
versus $N = 2, D = 8$  supergravity'', 
{\tt hep-th/0304221}.

\bibitem{horne}
J.~H.~Horne,
``Superspace versions of topological theories'',
{\em Nucl.Phys.}  {\bf B318} (1989) 22;

C.~Aragao de Carvalho and L.~Baulieu,
``Local BRST symmetry and superfield formulation of the 
Donaldson-Witten theory'', 
{\em Phys.Lett.}  {\bf B275} (1992) 323.

\bibitem{osvb} 
S.~Ouvry, R.~Stora and P.~van Baal,
``On the algebraic characterization of Witten's topological Yang-Mills
theory'',  
{\em Phys.Lett.}  {\bf B220} (1989) 159.


\bibitem{opetal}
M.~Blau and G.~Thompson, 
``Aspects of $N_T \geq 2$ topological gauge theories and D-branes'',
{\em Nucl.Phys.} {\bf B492} (1997) 545, 
{\tt hep-th/9612143}; 

B.~Geyer and D.~M\"ulsch,
``$N_T =4$ equivariant extension of the 3D topological model
of Blau and Thompson'',
{\em Nucl.Phys.} {\bf B616} (2001) 476, 
{\tt hep-th/0108042};

B.~Geyer and D.~M\"ulsch,
``Higher dimensional analogue of the 
Blau-Thompson model and $N_T =8, D=2$ Hodge-type cohomological 
gauge theories'',
{\em Nucl.Phys.} {\bf B662} (2003) 531, {\tt hep-th/0211061}; 

C.P.~Constantinidis, O.~Piguet and W.~Spalenza,
``Superspace gauge fixing of topological Yang-Mills theories'',
{\em European Phys. J.} (2004) to appear, 
{\tt hep-th/0310184}.

\bibitem{bcglp}  
J.L.~Boldo, C.P.~Constantinidis, F.~Gieres, 
M.~Lefran\c cois and O.~Piguet, 
``Observables in topological Yang-Mills theories'',
{\em Int.J.Mod.Phys.} {\bf A} (2003) to appear, 
{\tt hep-th/0303053};

J.L.~Boldo, C.P.~Constantinidis, F.~Gieres, 
M.~Lefran\c cois and O.~Piguet, 
``Topological Yang-Mills theories and their observables:
a superspace approach'', 
{\em Int.J.Mod.Phys.} {\bf A18} (2003) 2119, 
{\tt hep-th/0303084}.

\bibitem{kanno} 
H.~Kanno, ``Weil algebra structure and geometrical meaning of 
BRST transformation in topological quantum field theory",
{\em Z. Physik} {\bf C43} (1989) 477;

H.~Kanno, ``Observables in  topological Yang-Mills theory and the 
Gribov problem", 
{\em Lett.Math.Phys.} {\bf 19} (1990) 249.


\bibitem{yuwu}
F.~Yu and Y.~S.~Wu,
``BRST superspace formulation for world sheet 
topological gravity with gauged Lorentz 
and Weyl symmetries'',
{\em Class.Quant.Grav.} {\bf 6} (1989) L199;

%\bibitem{tahiri}
M.~Tahiri,
``Unconstrained superspace formalism 
of topological 2-d gravity'',
{\em Mod.Phys.Lett.} {\bf A10} (1995) 1949.

\bibitem{bb} 
L.~Baulieu and M.~Bellon,
``$p$-forms and supergravity: gauge symmetries in curved space'', 
{\em Nucl.Phys.} {\bf B266} (1986) 75. 

\bibitem{bertlmann}
R.~A. Bertlmann, {\em {Anomalies in quantum field theory}}
\newblock (Clarendon Press, Oxford 1996).


\bibitem{bf} 
C.P.~Constantinides, F.~Gieres, O.~Piguet
and M.S.~Sarandy, 
``On the symmetries of BF models and their relation with gravity'',
{\em JHEP} {\bf 0201} (2002) 017,
{\tt hep-th/0111273}.



\bibitem{exp-xi}
F.~Langouche, T.~Sch\"ucker and R.~Stora,
``Gravitational anomalies of the Adler-Bardeen type'', 
{\em Phys.Lett.}
{\bf B145} (1984) 342;

L.~Baulieu and J.~Thierry-Mieg,
``Algebraic structure of quantum gravity and the classification of the
gravitational anomalies'',
{\em Phys.Lett.} {\bf B145} (1984) 53. 

\bibitem{bbrt}
D.~Birmingham, M.~Blau, M.~Rakowski and G.~Thompson,
``Topological field theory'', 
{\em Phys.Rept.} {\bf 209} (1991) 129.

\bibitem{op}
O.~Piguet, ``Ghost equations and diffeomorphism invariant theories'',\\
{\em Class.Quant.Grav.} {\bf 17} (2000) 3799, 
{\tt hep-th/0005011}. 

\bibitem{ny}
H.T.~Nieh and M.L.~Yan,
``An identity in Riemann-Cartan geometry'', \\
{\em J.Math.Phys.}  {\bf 23} (1982) 373; 

H.T.~Nieh and M.L.~Yan,
``Quantized Dirac field in curved Riemann-Cartan background. 1. Symmetry
properties, Green's function'',
{\em Annals Phys.} {\bf 138} (1982) 237.

\bibitem{zanelli}
O.~Chand\'{\i}a and J.~Zanelli,
``Torsional topological invariants (and their relevance for real life)'',
in {\em Trends in Theoretical Physics},  
AIP Conference Proceedings Vol. 419, 
H.~Falomir, R.E.~Gamboa-Sarav\'{\i}, F.A.~Schaposnik, eds. 
(Amer. Inst. Phys., 1998), 
{\tt hep-th/9708138}; 

O.~Chand\'{\i}a and J.~Zanelli,
 ``Topological invariants, instantons and chiral ano\-maly on spaces with
torsion'', 
{\em Phys.Rev.}  {\bf D55}  (1997) 7580, 
{\tt hep-th/9702025}; 

A.~Mardones and J.~Zanelli,
``Lovelock-Cartan theory of gravity'', \\
{\em Class.Quant.Grav.}  {\bf 8}  (1991) 1545;  

H.Y.~Guo, K.~Wu and W.~Zhang,
``On torsion and Nieh-Yan form'', \\
{\em Commun.Theor.Phys.}  {\bf 32} (1999) 381,
{\tt hep-th/9805037}.


\bibitem{bbg} 
L.~Baulieu, M.~Bellon and R.~Grimm,
``BRS symmetry of supergravity in superspace and its projection 
to component formalism'', 
{\em Nucl.Phys.} {\bf B294} (1987) 279. 


\bibitem{witten-3d-grav} 
E.~Witten, 
``(2+1)-dimensional gravity as an exactly soluble system'',\\
{\em Nucl.Phys.} {\bf B311} (1988) 46; 

E.~Witten, 
``Topology changing amplitudes in (2+1)-dimensional gravity'',
{\em Nucl.Phys.} {\bf B323} (1989) 113.


\bibitem{birmrak}
D.~Birmingham and M.~Rakowski,
``Equivariance in topological gravity,''
{\em Phys.Lett.}  {\bf B289} (1992) 271.



\bibitem{eguhan}
T.~Eguchi and A.J.~Hanson, 
``Self-dual solutions to euclidean gravity'',
{\em Annals Phys.}  {\bf 120} (1979) 82.

\bibitem{brand}
A.~Brandhuber, O.~Moritsch, M.~W.~de Oliveira, O.~Piguet and M.~Schweda,
``A renormalized supersymmetry 
in the topological Yang-Mills field theory'',
{\em Nucl.Phys.} {\bf B431} (1994) 173, 
{\tt hep-th/9407105};

F.~Gieres, J.~Grimstrup, T.~Pisar and M.~Schweda, 
``Vector supersymmetry in topological field theories'',
{\em JHEP} {\bf 0006} (2000) 018, 
{\tt hep-th/0002167}.


\bibitem{ader}
J.~P.~Ader, F.~Gieres and Y.~Noirot,
``Gauged BRST symmetry and covariant gravitational anomalies'',
{\em Phys.Lett.} {\bf B256} (1991) 401.

\bibitem{klt}
M.~Knecht, S.~Lazzarini and F.~Thuillier,
``Shifting the Weyl anomaly to the chirally split 
diffeomorphism anomaly in
two dimensions'',
{\em Phys.Lett.} {\bf B251} (1990) 279.

\end{thebibliography}
\end{document}